\pdfoutput=1

\documentclass[11pt,twoside,a4paper,cmspaper,final,collab]{cms-tdr}
\def\svnVersion{440432P}\def\cmsCernNoTag{CERN-EP-2017-087}\def\cmsCernDate{\today}\def\cmsMessage{Published in the European Physical Journal C as \href{http://dx.doi.org/10.1140/epjc/s10052-017-5317-4}{\doi{10.1140/epjc/s10052-017-5317-4.}\\Author list corrected in this version.}}

\begin{document}\cmsNoteHeader{EXO-16-005}

\hyphenation{had-ron-i-za-tion}
\hyphenation{cal-or-i-me-ter}
\hyphenation{de-vices}
\newcommand{\ifpas}{\iftrue}
\RCS$Revision: 363198 $
\RCS$HeadURL: svn+ssh://svn.cern.ch/reps/tdr2/papers/EXO-16-005/trunk/EXO-16-005.tex $
\RCS$Id: EXO-16-005.tex 363198 2016-08-03 18:14:29Z ksung $
\newlength\cmsFigWidth
\ifthenelse{\boolean{cms@external}}{\setlength\cmsFigWidth{0.80\columnwidth}}{\setlength\cmsFigWidth{0.4\textwidth}}
\ifthenelse{\boolean{cms@external}}{\providecommand{\cmsLeft}{upper\xspace}}{\providecommand{\cmsLeft}{left\xspace}}
\ifthenelse{\boolean{cms@external}}{\providecommand{\cmsRight}{lower\xspace}}{\providecommand{\cmsRight}{right\xspace}}
\providecommand{\NA}{\ensuremath{\text{---}}\xspace}
\ifthenelse{\boolean{cms@external}}{\providecommand{\cmsTable}[1]{#1}}{\providecommand{\cmsTable}[1]{\resizebox{\textwidth}{!}{#1}}}
\newcommand{\Znunu}{\ensuremath{\Z(\nu\bar{\nu})+\text{jets}}}
\newcommand{\Zll}{\ensuremath{\Z(\ell\ell)}+\text{jets}}
\newcommand{\Zee}{\ensuremath{\Z(\Pe\Pe)}}
\newcommand{\Zmm}{\ensuremath{\Z(\PGm\PGm)}}
\newcommand{\Wlnu}{\ensuremath{\PW(\ell\nu)+\text{jets}}}
\newcommand{\Wjets}{\ensuremath{\PW+\text{jets}}}
\newcommand{\Zjets}{\ensuremath{\Z+\text{jets}}}
\newcommand{\WZjets}{\ensuremath{\text{W/Z}+\text{jets}}}
\newcommand{\mll}{\ensuremath{m_{\ell\ell}}}
\newcommand{\gqq}{\ensuremath{g_{\PQq\PQq}}}
\newcommand{\XX}{\ensuremath{\chi\overline{\chi}}\xspace}
\newcommand{\ttbb}{\ensuremath{\ttbar/\bbbar}\xspace}
\newcommand{\ttDM}{\ensuremath{\ttbar+\chi\overline{\chi}}\xspace}
\newcommand{\bbDM}{\ensuremath{\bbbar+\chi\overline{\chi}}}
\newcommand{\htmiss}{\ensuremath{{ H_{\mathrm{T}}^{\text{miss}}}}\xspace}
\newcommand{\ttMET}{\ensuremath{\ttbar+\ptmiss}\xspace}
\newcommand{\bbMET}{\ensuremath{\bbbar+\ptmiss}\xspace}
\newcommand{\mg}{\textsc{MG5}\_a\MCATNLO\xspace}
\newcommand{\ee}{\ensuremath{\Pe\Pe}\xspace}
\newcommand{\emu}{\ensuremath{\Pe\mu}\xspace}
\newcommand{\mumu}{\ensuremath{\PGm\PGm}\xspace}
\newcommand{\eorm}{\ensuremath{\Pe,\PGm}\xspace}
\newcommand{\tteeMET}{\ensuremath{\ttbar(\ee)+\ptmiss}\xspace}
\newcommand{\ttemMET}{\ensuremath{\ttbar(\emu)+\ptmiss}\xspace}
\newcommand{\ttmmMET}{\ensuremath{\ttbar(\mumu)+\ptmiss}\xspace}
\newcommand{\tteormMET}{\ensuremath{\ttbar(\Pe,\PGm)+\ptmiss}\xspace}
\newcommand{\ttoneRTTMET}{\ensuremath{\ttbar(0,1\mathrm{RTT})+\ptmiss}\xspace}
\newcommand{\tttwoRTTMET}{\ensuremath{\ttbar(2\text{RTT})+\ptmiss}\xspace}
\newcommand{\mtop}{\ensuremath{m_{\text{top}}}\xspace}
\newcommand{\mt}{\ensuremath{M_{\mathrm{T}}}\xspace}
\newcommand{\mwt}{\ensuremath{M^{\PW}_{\mathrm{T2}}}\xspace}
\newcommand{\mindphi}{\ensuremath{\min\Delta\phi(\ptvecjet,\ptvecmiss)}\xspace}
\newcommand{\mindphill}{\ensuremath{\min\Delta\phi(\ptvec^{\ell\ell},\ptvecmiss)}\xspace}
\newcommand{\dphill}{\ensuremath{\Delta\phi(\ptvec^{\ell\ell},\ptvecmiss)}\xspace}
\newcommand{\semileptonic}{\ensuremath{\ell+\text{jets}}\xspace}
\newcommand{\gq}{\ensuremath{g_{\PQq}}\xspace}
\newcommand{\muF}{\ensuremath{\mu_{\text{F}}}\xspace}
\newcommand{\muR}{\ensuremath{\mu_{\text{R}}}\xspace}
\newcommand{\ptvecjet}{\ensuremath{\ptvec^{\kern1pt\mathrm{jet_{i}}}}\xspace}

\cmsNoteHeader{EXO-16-005}
\title{Search for dark matter produced in association with heavy-flavor quark pairs in proton-proton collisions at $\sqrt{s}= 13\TeV$}
\titlerunning{Search for dark matter produced with  heavy-flavor quark pairs  at $\sqrt{s}= 13\TeV$}

\date{\today}

\abstract{
A search is presented for an excess of events with heavy-flavor quark pairs ($\ttbar$ and $\bbbar$) and a large imbalance in transverse momentum in data from proton-proton collisions at a center-of-mass energy of 13\TeV. The data correspond to an integrated luminosity of 2.2\fbinv collected with the CMS detector at the CERN LHC.  No deviations are observed with respect to standard model predictions.  The results are used in the first interpretation of dark matter production in $\ttbar$ and $\bbbar$ final states in a simplified model.  This analysis is also the first to perform a statistical combination of searches for dark matter produced with different heavy-flavor final states.  The combination provides exclusions that are stronger than those achieved with individual heavy-flavor final states.
}

\hypersetup{%
pdfauthor={CMS Collaboration},%
pdftitle={Search for dark matter produced in association with heavy-flavor quarks in proton-proton collisions at sqrt(s)=13 TeV},%
pdfsubject={CMS},%
pdfkeywords={CMS, dark matter, physics}}

\maketitle

\section{Introduction}\label{intro}
Astrophysical and cosmological observations~\cite{Bertone:2004pz,Feng:2010gw,Porter:2011nv} provide strong support for the existence of dark matter (DM), which could originate from physics beyond the standard model (BSM).  In a large class of BSM models, DM consists of stable, weakly-interacting massive particles (WIMPs).  In collider experiments, WIMPs ($\chi$) could be pair-produced through the exchange of new mediating fields that couple to DM and to standard model (SM) particles.  Following their production, the WIMPs would escape detection, thereby creating an imbalance of transverse momentum (missing transverse momentum, $\ptmiss$) in the event.

\begin{figure}[h!tb]
\centering
\includegraphics[width=\cmsFigWidth]{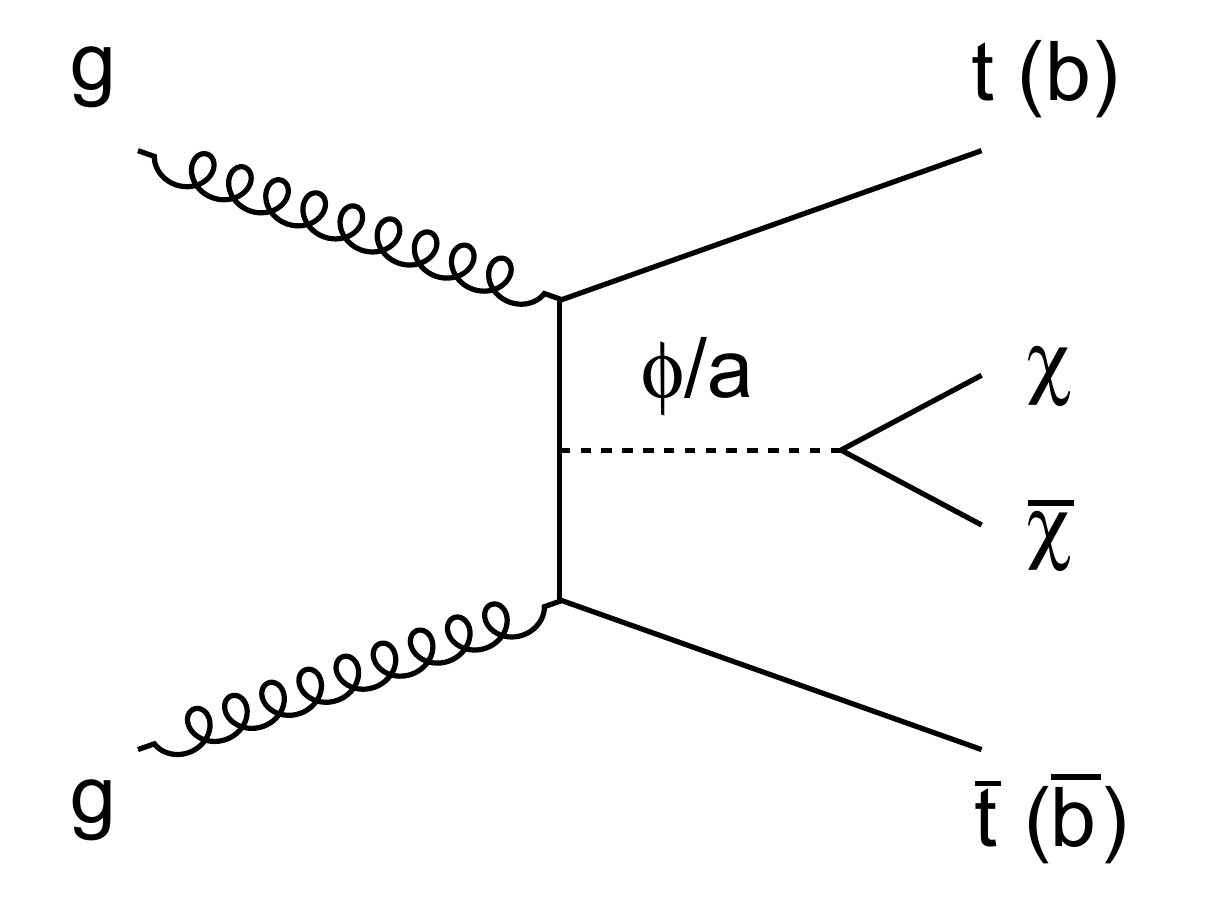}
\caption{A leading order Feynman diagram describing the production of a pair of DM particles ($\chi$) with heavy-flavor (top or bottom) quark pairs via scalar ($\phi$) or pseudoscalar ($\mathrm{a}$) mediators.}
\label{fig:feynman}
\end{figure}

If the new physics associated with DM respects the principle of minimal flavor violation~\cite{D'Ambrosio:2002ex,Isidori:2012ts}, the interactions of spin-0 mediators retain the Yukawa structure of the SM.  This principle is motivated by the apparent lack of new flavor physics at the electroweak (EWK) scale.  Because only the top quark has a Yukawa coupling of order unity, WIMP DM couples preferentially to the heavy top quark in models with minimal flavor violation.  In high energy proton-proton collisions, this coupling leads to the production of $\ttDM$ at lowest-order via a scalar ($\phi$) or pseudoscalar (a) mediator (Fig.~\ref{fig:feynman}), and to the production of so-called mono-X final states through a top quark loop~\cite{Haisch:2012kf,Lin:2013sca,Buckley:2014fba,Haisch:2015ioa,Arina:2016cqj,Aad:2015zva,Khachatryan:2016mdm,Aaboud:2016tnv,Sirunyan:2017hci}.  At the CERN Large Hadron Collider (LHC), the $\ttDM$ process can be probed directly via the $\ttMET$ and $\bbMET$ signatures.  The $\bbMET$ signature provides additional sensitivity to the $\bbDM$ process for models in which mediator couplings to up-type quarks are suppressed, as can be the case in Type-II two Higgs doublet models~\cite{Branco:2011iw}.

This paper describes a search for DM produced with a $\ttbar$ or $\bbbar$ pair in pp collisions at $\sqrt{s}=13\TeV$ with the CMS experiment at the LHC.  A potential DM signal is extracted from simultaneous fits to the $\ptmiss$ distributions in the $\bbMET$ and $\ttMET$ search channels.  Data from control regions enriched in SM $\ttbar$, $\Wjets$, and $\Zjets$ processes are included in the fits, to constrain the major backgrounds.  The top quark nearly always decays to a W boson and a b quark.  The W boson subsequently decays leptonically (to charged leptons and neutrinos) or hadronically (to quark pairs).  The dileptonic, lepton($\ell$)+jets, and all-hadronic $\ttbar$ final states consist, respectively, of events in which both, either, or neither of the W bosons decay leptonically.  Each of these primary $\ttbar$ final states are explored.

Previous LHC searches for DM produced with heavy-flavor quark pairs were interpreted using effective field theories that parameterize the DM-SM coupling in terms of an interaction scale $M_{*}$~\cite{Beltran:2010ww,Goodman:2010ku,Cheung:2010zf}.  An earlier search by the CMS Collaboration investigated the \semileptonic $\ttbar$ final state using $19.7\fbinv$ of data collected at $\sqrt{s} = 8\TeV$~\cite{Khachatryan:2015nua}.  That search excluded values of $M_{*}$ below 118\GeV, assuming $m_{\chi} = 100\GeV$.  The ATLAS Collaboration performed a similar search separately for the all-hadronic and \semileptonic $\ttbar$ final states and obtained comparable limits on $M_{*}$~\cite{Aad:2014vea}.  More recently, the limitations of effective field theory interpretations of DM production at the LHC has led to the development of simplified models that remain valid when the mediating particle is produced on-shell~\cite{Abercrombie:2015wmb}.  This analysis adopts the simplified model framework to provide the first interpretation of heavy-flavor search results in terms of the decays of spin-0 mediators with scalar or pseudoscalar couplings.  This paper also reports the first statistical combination of dileptonic ($\ee$, $\emu$, $\mumu$), \semileptonic (\Pe, \PGm), and all-hadronic $\ttDM$ searches, as well as the first combination of $\ttDM$ and $\bbDM$ search results.

The paper is organized as follows.  Section~\ref{sec:detector} reviews the properties of the CMS detector and the particle reconstruction algorithms used in the analysis.  Section~\ref{sec:simulation} describes the modeling of $\ttDM$ and $\bbDM$ signal and SM background events, and Section~\ref{selection} provides the selections applied to data and simulation.  Section~\ref{sec:extract} discusses the techniques used to extract a potential DM signal in the $\ttMET$ and $\bbMET$ search channels.  Section~\ref{sec:syst} describes the systematic uncertainties considered in the analysis.  The results of the search and their interpretation within a simplified DM framework are presented in Section~\ref{sec:results}.  Section~\ref{sec:summary} concludes with a summary of the results.

\section{CMS detector and event reconstruction}\label{sec:detector}
The CMS detector~\cite{CMS} is a multipurpose apparatus optimized for the study high transverse momentum ($\pt$) physics processes in pp and heavy ion collisions. A superconducting solenoid surrounds the central region, providing a magnetic field of 3.8\unit{T} parallel to the beam direction. Charged particle trajectories are measured using the silicon pixel and strip trackers, which cover the pseudorapidity region of $\abs{\eta}< 2.5$. A lead tungstate crystal electromagnetic calorimeter (ECAL) and a brass and scintillator hadron calorimeter (HCAL) surround the tracking volume, and cover the region with $\abs{\eta}< 3$.  Each calorimeter is composed of a barrel and two endcap sections.  A steel and quartz-fiber Cherenkov forward hadron calorimeter extends the coverage to $\abs{\eta}< 5$. The muon system consists of gas-ionization detectors embedded in the steel flux return yoke outside the solenoid, and covers the region of $\abs{\eta}< 2.4$. The first level of the CMS trigger system is composed of special hardware processors that select the most interesting events in less than 4\mus using information from the calorimeters and muon detectors.  This system reduces the event rate from 40\unit{MHz} to approximately 100\unit{kHz}.  The high-level trigger processor farm performs a coarse reconstruction of events selected by the first-level trigger, and applies additional selections to reduce the event rate to less than 1\unit{kHz} for storage.

Event reconstruction is based on the CMS Particle Flow (PF) algorithm~\cite{PF1,PF2}, which combines information from all CMS subdetectors to identify and reconstruct the individual particles emerging from a collision: electrons, muons, photons, and charged and neutral hadrons.  Interaction vertices are reconstructed using the deterministic annealing algorithm~\cite{Chatrchyan:2014fea}.  The primary vertex is selected as that with the largest sum of $\pt^{2}$ of its associated charged particles. Events are required to have a primary vertex that is consistent with being in the luminous region.

Jets are reconstructed by clustering PF candidates using the anti-\kt algorithm~\cite{Cacciari:2008gp,Cacciari:2011ma} with a distance parameter of 0.4. Corrections based on jet area are applied to remove the energy from additional collisions in the same or neighboring bunch crossing (pileup)~\cite{Cacciari:2008gn}.  Energy scale calibrations determined from the comparison of simulation and data are then applied to correct the four momenta of the jets~\cite{2011JInst...611002C}.  Jets are required to have $\pt > 30\GeV$, $\abs{\eta}< 2.4$, and to satisfy a loose set of identification criteria designed to reject events arising from spurious detector and reconstruction effects.

The combined secondary vertex b tagging algorithm (CSVv2) is used to identify jets originating from the hadronization of bottom quarks~\cite{1748-0221-8-04-P04013,btag}. Jets are considered to be b-tagged if the CSVv2 discriminant for that jet passes a requirement that roughly corresponds to efficiencies of 70\% to tag bottom quark jets, 20\% to mistag charm quark jets, and 1\% to misidentify light-flavor jets as b jets.  Efficiency scale factors in the range of 0.92--0.98, varying with jet \pt, are applied to simulated events in order to reproduce the b tagging performance for bottom and charm quark jets observed in data.  A scale factor of 1.14 is applied to simulation to reproduce the measured mistag rate for light-flavor quark and gluon jets.

The $\ptmiss$ variable is initially calculated as the magnitude of the vector sum of the \pt of all PF particles.  This quantity is adjusted by applying jet energy scale corrections.  Detector noise, inactive calorimeter cells, and cosmic rays can give rise to events with severely miscalculated $\ptmiss$.  Such events are removed via a set of quality filters that take into account the timing and distribution of signals from the calorimeters, missed tracker hits, and global characteristics of the event topology.

Electron candidates are reconstructed by combining tracking information with energy depositions in the ECAL \cite{Khachatryan:2015hwa}. The energy of the ECAL clusters is required to be compatible with the momentum of the associated electron track. Muon candidates are reconstructed by combining tracks from the inner silicon tracker and the outer muon system~\cite{Chatrchyan:2012xi}. Tracks associated with muon candidates must be consistent with a muon originating from the primary vertex, and must satisfy a set of quality criteria~\cite{Chatrchyan:2012xi}.  Electrons and muons are selected with $\pt > 30\GeV$ and $\abs{\eta}< 2.1$ for consistency with the coverage of the single-lepton triggers, and are required to be isolated from hadronic activity, to reject hadrons misidentified as leptons. Relative isolation is defined as the scalar \pt sum of PF candidates within a $\Delta{R} = \sqrt{\smash[b]{\eta^{2} + \phi^{2}}}$ cone of radius 0.4 or 0.3 centered on electrons or muons, respectively, divided by the lepton $\pt$.  Relative isolation is nominally required to be less than 0.035 (0.065) for electrons in the barrel (endcap), respectively, and less than 0.15 for muons.  Identification requirements, based on hit information in the tracker and muon systems, and on energy depositions in the calorimeters, are imposed to ensure that candidate leptons are well-measured.  These restrictive isolation and identification criteria are used to select events from the dileptonic $\ttbar$, \semileptonic $\ttbar$, $\Wlnu$, and $\Zll$ processes.

The efficiencies of the requirements for electrons (muons) with $\pt > 30\GeV$ range from 52 to 83\% (91 to 96\%), for increasing lepton $\pt$.  Less restrictive lepton isolation and identification requirements are used to reject events containing additional leptons with $\pt > 10\GeV$.  Efficiencies for these requirements range from 66 to 96\% for electrons and 73 to 99\% for muons, for increasing lepton $\pt$.  Electron and muon selection efficiency scale factors are applied in simulation to match the efficiencies measured in data using the tag-and-probe procedure~\cite{CMS:2011aa}.  Averaged over lepton $\pt$, the electron and muon efficiency scale factors for the more restrictive selection requirements are 98 and 99\%, respectively.

The ``resolved top tagger'' (RTT) is a multivariate discriminant that uses jet properties and kinematics to identify top quarks that decay into three resolved jets.  The input observables are the values of the quark/gluon discriminant~\cite{JME-14-002}, which combines track multiplicity, jet shape, and fragmentation information for each jet, values of the b tagging discriminants, and the opening angles between the candidate b jet and the two jets from the candidate W boson.  Within each jet triplet, the b candidate is considered to be the jet with the largest value of the b tagging discriminant.  The RTT discriminant also utilizes the $\chi^{2}$ value of a simultaneous kinematic fit to the top quark and W boson masses~\cite{D'Hondt:926540}.  The fit attempts to satisfy the mass constraints by allowing the jet momenta and energies to vary within their measured resolutions.  The RTT is implemented as a boosted decision tree using the TMVA framework~\cite{Hocker:2007ht}, and is trained on simulated \semileptonic $\ttbar$ events using correct (incorrect) jet combinations as signal (background).

\begin{figure}[htb]\centering
 \includegraphics[width=0.49\textwidth]{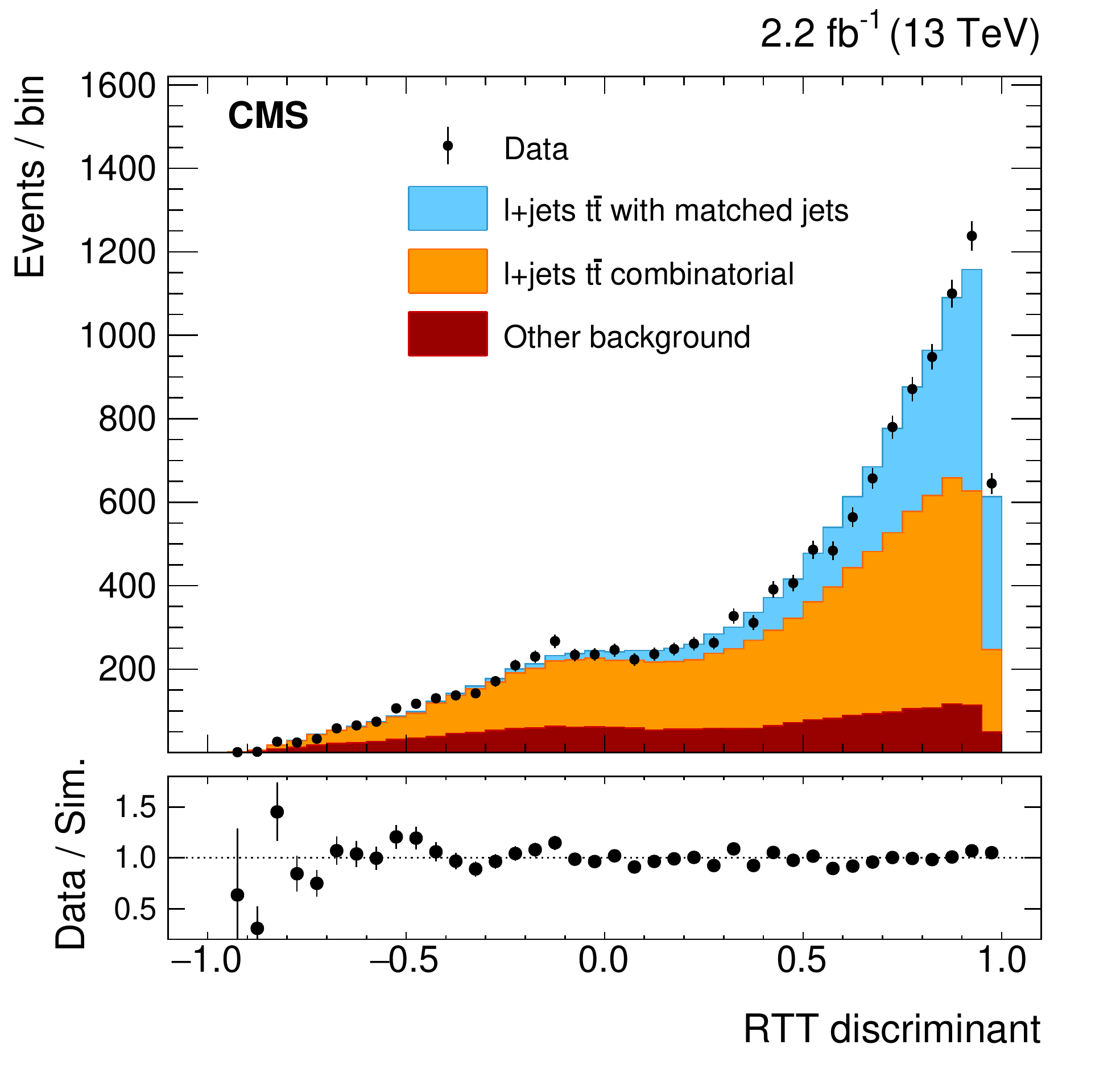}
 \caption{The distribution of the RTT discriminant in data enriched in \semileptonic $\ttbar$ events. Simulated \semileptonic $\ttbar$ events in which jets from the all-hadronic top quark decay are correctly chosen are labeled ``$\ttbar(1\ell)$ with matched jets''.  Simulated \semileptonic $\ttbar$ events in which an incorrect combination of jets is chosen are labeled ``$\ttbar(1\ell)$ combinatorial''.  Events from processes that do not contain a hadronically-decaying top quark, such as dileptonic $\ttbar$, are labeled ``other background''.  The uncertainties shown in the ratios of data to simulation are statistical only.  Jet triplets in the all-hadronic $\ttMET$ search are considered to be top quark tagged if their RTT discriminant value is larger than zero.}
 \label{fig:resolved}
\end{figure}

The performance of the RTT discriminant is characterized with data enriched in SM \semileptonic $\ttbar$ events containing four or more jets.  At least one of these jets is required to be b-tagged.  The output discriminant for these events is plotted in Fig.~\ref{fig:resolved}.  Each entry in the plot corresponds to the jet triplet with the highest RTT score in the event.  Data are modeled using simulated \semileptonic $\ttbar$ signal events, and simulated events for each of the primary backgrounds (dileptonic $\ttbar$, $\Wjets$, single t).  The simulation is split into three classes that correspond to correctly tagged jet triplets and the two possibilities for mistagging, as explained below.  Simulation describes the data well.  A jet triplet is considered as a tagged top quark decay when the RTT discriminant value is greater than zero.

There are three efficiencies associated with the RTT selection, which correspond to the three classes of events in Fig.~\ref{fig:resolved}: \semileptonic $\ttbar$ events in which the hadronically-decaying top quark is correctly identified (``$\ttbar(1\ell)$ matched''), \semileptonic $\ttbar$ events in which an incorrect combination of jets is tagged (``$\ttbar(1\ell)$ combinatorial''), and events with no hadronically-decaying top quarks that contain a mistagged jet triplet (``other background'').  Dileptonic $\ttbar$ events are used to extract the nonhadronic mistag rate in data.  Then, \semileptonic $\ttbar$ events are used to extract the tagging and mistagging efficiencies for hadronically-decaying top quarks through a fit to the trijet mass distribution.  Mass templates obtained from simulation are associated with each efficiency term in the fit.  The efficiency of the RTT $>0$ selection for events determined to be $\ttbar(1\ell)$ matched, $\ttbar(1\ell)$ combinatorial, or other background are $0.97 \pm 0.03$, $0.80 \pm 0.05$, and $0.69 \pm 0.02$, respectively.  Corresponding data-to-simulation scale factors are found to be consistent with unity.

The $\bbMET$ search includes vetoes on hadronically-decaying $\tau$ leptons, which are reconstructed from PF candidates using the ``hadron plus strips'' algorithm~\cite{Chatrchyan:2012zz}.  The algorithm combines one or three charged pions with up to two neutral pions.  Neutral pions are reconstructed by the PF algorithm from the photons that arise from $\pi^0\to\gamma\gamma$ decay.  Photons are reconstructed from ECAL energy clusters, which are corrected to recover the energy deposited by photon conversions and bremsstrahlung.  Photons are identified and distinguished from jets and electrons using cut-based criteria that include the isolation and transverse shape of the ECAL deposit, and the ratio of HCAL/ECAL energies in a region surrounding the candidate photon.

\section{Modeling and simulation}\label{sec:simulation}
The associated production of DM and heavy-flavor quark pairs provides rich detector signatures that include significant $\ptmiss$ accompanied by high-$\pt$ jets, bottom quarks, and leptons.  The largest backgrounds in the $\ttMET$ and $\bbMET$ searches are SM $\ttbar$ events, inclusive W boson production in which the W decays leptonically ($\Wlnu$), and inclusive Z boson production in which the Z decays to neutrinos ($\Znunu$). Simulated events are used throughout the analysis to determine signal and background expectations.  Where possible, corrections determined from data are applied to the simulations.

Monte Carlo (MC) samples of SM $\ttbar$ and single t backgrounds are generated at next-to-leading order (NLO) in quantum chromodynamics (QCD) using {\POWHEG v2} and {\POWHEG v1}~\cite{Nason:2004rx,Frixione:2007vw,Alioli:2010xd}, respectively.  As with all MC generators subsequently described, {\POWHEG} is interfaced with {\PYTHIA 8.205}~\cite{Sjostrand:2007gs} for parton showering using the {CUETP8M1} tune~\cite{Khachatryan:2015pea}.  Samples of $\Zjets$, $\Wjets$, and QCD multijet events are produced at leading order (LO) using \mg v2.2.2~\cite{Alwall:2014hca} with the MLM prescription~\cite{Mangano:2006rw} for matching jets from the matrix element calculation to the parton shower description. The $\Wjets$ and $\Zjets$ samples are corrected using EWK and QCD NLO/LO K-factors calculated with \mg as functions of the generated boson $\pt$.  The simulation of $\ttbar+\gamma$, $\ttbar+\PW$, and $\ttbar+\Z$ events makes use of NLO matrix element calculations implemented in \mg{}, and the FxFx~\cite{Frederix:2012ps} prescription to merge multileg processes. Diboson processes (WW, WZ, and ZZ) are generated at NLO using either \mg or {\POWHEG v2}.

The signal processes are simulated using simplified models that were developed in the LHC Dark Matter Forum (DMF)~\cite{Abercrombie:2015wmb}.  The DM particles $\chi$ are assumed to be Dirac fermions, and the mediators are spin-0 particles with scalar ($\phi$) or pseudoscalar ($\mathrm{a}$) couplings.  The coupling strength of the mediator to SM fermions is assumed to be $\gqq = \gq y_{\PQq}$ where: $y_{\PQq}=\sqrt{2}m_{\PQq}/v$ is the SM Yukawa coupling, $m_{\PQq}$ is the quark mass, and $v = 246\GeV$ is the Higgs field vacuum expectation value.  As per the recommendations of the LHC Dark Matter Working Group~\cite{Boveia:2016mrp}, $\gq$ is taken to be flavor universal and equal to 1.  Likewise, the coupling strength of the mediator to DM, $g_{\chi}$, is set to 1 and is independent of the DM mass.  The LHC DMF spin-0 models do not account for mixing between the $\phi$ scalar and the SM Higgs boson~\cite{Bauer:2016gys}.  As is discussed in~\cite{Abercrombie:2015wmb}, the $\ptmiss$ spectra of both the scalar and pseudoscalar mediated processes broaden with increasing mediator mass.  For $m_{\phi/\mathrm{a}}$ larger than twice the top quark mass (\mtop), the $\ptmiss$ distributions of the scalar and pseudoscalar processes are essentially identical.  As $m_{\phi/\mathrm{a}}$ decreases below $2\mtop$, the $\ptmiss$ spectra of the two processes increasingly differ, with the distribution of the scalar process peaking at lower $\ptmiss$ values~\cite{Boudjema:2015nda,Backovic:2015soa}. For all mediator masses, the total cross section of the scalar process is larger than that of the pseudoscalar equivalent~\cite{Backovic:2015soa}. This analysis focuses on the $m_{\chi} = 1\GeV$ LHC DMF benchmark point, which provides a convenient signal reference for both low and high mass mediators.

The $\ttDM$ and $\bbDM$ signals are generated at LO in QCD using \mg with up to one additional jet in the final state.  Jets from the matrix element calculations are matched to the parton shower descriptions using the MLM prescription.  Angular correlations in the decays of the top quarks are included using \textsc{MadSpin} v2.2.2~\cite{Artoisenet:2012st}.  Minimum decay widths are assumed for the mediators, and are calculated from the partial width formulas given in Ref.~\cite{PhysRevD.91.055009}.  This calculation assumes that the spin-0 mediators couple only to SM quarks and the DM fermion $\chi$.  Simulated signal samples are produced for a DM mass of $m_{\chi} = 1\GeV$ and for mediator masses in the range of 10--500\GeV.  The relative width of the scalar (pseudoscalar) mediator varies between 4-6\% (4-8\%) for this mediator mass range.  The predicted rates of the $\bbDM$ process, which is generated in the 4-flavor scheme, are adjusted to match the cross sections calculated in the 5-flavor scheme~\cite{Maltoni:2012pa,Abercrombie:2015wmb}.

All samples generated at LO and NLO use corresponding NNPDF3.0~\cite{Ball:2014uwa} parton distribution function (PDF) sets.  All signal and background samples are processed using a detailed simulation of the CMS detector based on {\GEANTfour}~\cite{Agostinelli:2002hh}. The samples are reweighted to account for the distribution of pileup observed in data.

\section{Event selection}\label{selection}
Signal events are expected to exhibit both large $\ptmiss$ from the production of two noninteracting DM particles and event topologies consistent with the presence of top quarks or b quark jets.  Data are therefore collected using triggers that select events containing large $\ptmiss$ or high-$\pt$ leptons.  Data for the dileptonic and \semileptonic $\ttMET$ searches are obtained using single-lepton triggers that require an electron (muon) with $\pt \ge 27 ~(20)\GeV$.  These trigger selections are more than $90\%$ efficient for PF-reconstructed electrons and muons that satisfy the $\pt$, identification, and isolation requirements imposed.  The trigger used for the $\bbMET$ and all-hadronic $\ttMET$ searches selects events based on the amount of $\ptmiss$ and $\htmiss$ reconstructed using a coarse version of the PF algorithm.  The $\htmiss$ variable is defined as the magnitude of the vector sum of the \pt of all jets in the event with $\pt > 20\GeV$, $|\eta|<5.0$.  Jets reconstructed from detector noise are removed in the $\htmiss$ calculation by additionally requiring neutral hadron energy fractions of less than 0.9.  The $\ptmiss$ and $\htmiss$ requirements for this trigger are 120\GeV.  The trigger is nearly 100\% efficient for events that satisfy subsequent selections based on fully-reconstructed PF $\ptmiss$.

Additional selections, described in Section~\ref{sec:selection_SR} and summarized in Table~\ref{tab:selections_SR}, are applied to define eight independent regions of data that are sensitive to DM signals: two $\bbMET$, one \semileptonic $\ttMET$, three dileptonic $\ttMET$, and two all-hadronic $\ttMET$ regions. Control regions (CRs) enriched in various background processes are also defined and are used to improve background estimates in the aforementioned signal regions (SRs).  In the CRs, individual signal selection requirements are inverted to enhance background yields and to prevent event overlaps with the SRs.  Collectively, the SRs and CRs associated with the individual $\ttDM$ and $\bbDM$ production and decay modes are referred to as ``channels''.  The $\bbDM$ channel and the three $\ttDM$ channels are used in simultaneous $\ptmiss$ fits (described in Section~\ref{sec:extract}) to extract a potential DM signal.  The fits allow the background-enriched CRs to constrain the contributions of SM $\ttbar$, $\Wjets$, and $\Zjets$ processes within the CRs and SRs of each channel.  The selections used to define the SRs and CRs are described in Sections~\ref{sec:selection_SR} and~\ref{sec:selection_CR}, respectively.  Tables~\ref{tab:selections_SR} and~\ref{tab:selections_CR} briefly summarize these selections.  Table~\ref{tab:selections_CR} defines a CR labeling scheme that is extensively used in subsequent sections.

\begin{table*}[htb]
  \topcaption{Overview of the selection criteria used to define the eight $\ttMET$ and $\bbMET$ signal regions.  The signal region selections (including the definitions of the variables $\mt$ and $\mwt$) are described in detail in Section~\ref{sec:selection_SR}.  Vetoes are applied in the dileptonic $\ttMET$ signal region to remove overlaps with the \semileptonic $\ttMET$ and $\bbMET$ control regions.  These control regions are summarized in Table~\ref{tab:selections_CR} and discussed in Section~\ref{sec:selection_CR}.}
  \label{tab:selections_SR}
  \centering
    \def\arraystretch{1.3}
     \cmsTable{ \begin{tabular}{c||c|c|c|c|l}
        \hline
        \multirow{3}{*}{Signal regions} &  \multirow{3}{*}{Leptons} & \multirow{3}{*}{Jets} & \multirow{3}{*}{b jets} & \multirow{3}{*}{$\ptmiss$}  & \multirow{3}{*}{Other selections} \\
                                        &            &        &          &           & \\
                                        &            &        &          &           & \\
        \hline
        \hline
        \multirow{6}{*}{Dileptonic $\ttMET$}     &  \multirow{2}{*}{$\ee$}   & \multirow{6}{*}{${\geq}2$}   & \multirow{6}{*}{${\geq}1$} & \multirow{6}{*}{${\geq}50\GeV$} & $\mindphill>1.2\unit{rad}$ \\
         &   &  &   &  & $m_{\ell\ell} > 20\GeV$\\
        \cline{2-2} &  \multirow{2}{*}{$\emu$}   &  &   &  & $\abs{m_{\ee,\mu\mu} - m_{Z}} > 15\GeV$  \\
         &   &  &   &  & Dileptonic $\ttbar$ control region veto  \\
        \cline{2-2} &  \multirow{2}{*}{$\mumu$}   &  &   &  & $\Zjets$ control region veto   \\
         &     &  &   &  & ~  \\
        \hline
        \hline
        \multirow{3}{*}{\semileptonic $\ttMET$} &  \multirow{3}{*}{$\Pe$ or $\PGm$}   & \multirow{3}{*}{$\geq$3}   & \multirow{3}{*}{$\geq$1}  & \multirow{3}{*}{${\geq} 160\GeV$}   & $\mt  > 160\GeV$ \\
        & & & & & $\mwt > 200\GeV$\\
        & & & & & $\mindphi>1.2\unit{rad}$\\
        \hline
        \hline
        \multirow{4}{*}{All-hadronic $\ttMET$}  &  \multirow{4}{*}{0} & \multirow{2}{*}{${\geq}4$}   & \multirow{2}{*}{${\geq}2$}  & \multirow{4}{*}{${\geq}200\GeV$}   & 0,1RTT\\
                                                      &                     &                             &                            &                                      & $\mindphi>1.0\unit{rad}$ \\
        \cline{3-4}\cline{6-6}
                                                      &                     & \multirow{2}{*}{${\geq}6$}   & \multirow{2}{*}{${\geq}1$}  &                                      & 2RTT \\
                                                      &                     &                             &                            &                                      & $\mindphi>0.4\unit{rad}$ \\
        \hline
        \hline
        \multirow{2}{*}{$\bbMET$}  &  \multirow{2}{*}{0} & 1 or 2   & 1  & \multirow{2}{*}{${\geq}200\GeV$}   & \multirow{2}{*}{$\mindphi>0.5\unit{rad}$}\\
        \cline{3-4}
                       & & 2 or 3   & 2  & & \\
        \hline
        \end{tabular}}
\end{table*}

\begin{table*}[phtb]
  \topcaption{Overview of the selection criteria used to define the background control regions associated with the $\ttMET$ and $\bbMET$ signal regions.  The control region selections are described in detail in Section~\ref{sec:selection_CR}.}
  \label{tab:selections_CR}
  \centering
    \def\arraystretch{1.3}
     \resizebox{\textwidth}{!}{
      \begin{tabular}{c|c|c|c|c|c|c|l}
        \hline
        & & & & & & & \\
       {Label} & {Associated signal region(s)} &  {Dominant background} & {Leptons} & {Jets} & {b jets} & $\ptmiss$ & {Additional or modified selections} \\
        & & & & & & & \\
        \hline
        \hline
        \multirow{2}{*}{slA} & \multirow{3}{*}{\semileptonic $\ttMET$ }  & \multirow{2}{*}{Dileptonic $\ttMET$}   &  \multirow{2}{*}{$\ee,\emu,\mumu$}       & \multirow{3}{*}{$\geq 3$}  & \multirow{2}{*}{${\geq}1$} & \multirow{3}{*}{$\geq 160\GeV$}   & No selection on $\mt, \mwt, \mindphi$\\
         &                                                          &                      &                                                  &                                          &                            &                                                                 &  bbC/bbD/bbE/bbH/bbI/bbJ control region veto\\
        \cline{1-1}\cline{3-4}\cline{6-6}\cline{8-8}
        slB  &  & $\Wjets$                         & $\Pe$ or $\PGm$     &                              &             0       &                                      & No selection on $\mwt,\mindphi$ \\
        \hline
         hadA & \multirow{5}{*}{Hadronic $\ttMET$,~~ 0,1RTT } & \semileptonic $\ttMET$   &  $\Pe$ or $\PGm$       & \multirow{5}{*}{$\geq 4$}  & ${\geq}2$ & \multirow{8}{*}{${\geq}200\GeV$}   & $\mt < 160\GeV$, 0,1RTT\\
        \cline{1-1}\cline{3-4}\cline{6-6}\cline{8-8}
                                                               hadB  &  & $\WZjets$                         &  0                  &                             & 0       &                                      & 0,1RTT \\
        \cline{1-1}\cline{3-4}\cline{6-6}\cline{8-8}
                                                               hadC  & & $\Wjets$                          &  $\Pe$ or $\PGm$       &                             & 0       &                                      & No selection on $\mt < 160\GeV$, $\mindphi$, 0,1RTT\\
        \cline{1-1}\cline{3-4}\cline{6-6}\cline{8-8}
                                                              \multirow{2}{*}{hadD} &    & \multirow{2}{*}{$\Zjets$}         &  \multirow{2}{*}{$\ee$ or $\mumu$}   &            & \multirow{2}{*}{0} &                           & $60 < m_{\ell\ell} < 120\GeV$ \\
        \cline{2-2}
        & \multirow{4}{*}{Hadronic $\ttMET$,~~ 2RTT } & & & & & & No selection on $\mindphi$  \\
        \cline{1-1}\cline{3-6}\cline{8-8}
                                                                 hadE & & \semileptonic $\ttMET$   &  $\Pe$ or $\PGm$       & \multirow{3}{*}{${\geq}6$}  & ${\geq}1$ &    & $\mt < 160\GeV$, $\geq 2\text{RTT}$\\
        \cline{1-1}\cline{3-4}\cline{6-6}\cline{8-8}
                                                                 hadF & & $\WZjets$                         &  0                  &                             & 0       &                                      & $\geq 2\text{RTT}$ \\
        \cline{1-1}\cline{3-4}\cline{6-6}\cline{8-8}
                                                                 hadG & & $\Wjets$                          &  $\Pe$ or $\PGm$       &                             & 0       &                                      & No selection on $\mt < 160\GeV$, $\mindphi$  , $\geq 2\text{RTT}$\\
        \hline
        \hline
        bbA & \multirow{5}{*}{$\bbMET$, ~~1~b tag }  & $\Wjets$   &  $\Pe$       & \multirow{5}{*}{1 or 2}  & \multirow{5}{*}{1} & \multirow{10}{*}{${\geq}200\GeV$}   & $50 < \mt < 160\GeV$\\
        \cline{1-1}\cline{4-4}
                                                        bbB & & \semileptonic $\ttbar$   & $\mu$     &                              &                    &                                      & No selection on $\mindphi$ \\
        \cline{1-1}\cline{3-4}\cline{8-8}
                                                        bbC & & \multirow{2}{*}{$\Zjets$}               & $\ee$      &                              &                    &                                      & $70 < m_{\ell\ell} < 110\GeV$\\
        \cline{1-1}\cline{4-4}
                                                        bbD & &                                         & $\mumu$  &                              &                    &                                      & No selection on $\mindphi$ \\
        \cline{1-1}\cline{3-4}\cline{8-8}
        bbE & & Dileptonic $\ttbar$                                & $\emu$    &                              &                    &                                      & No selection on $\mindphi$ \\ \cline{1-6}\cline{8-8}
        bbF & \multirow{5}{*}{$\bbMET$, ~~2~b tag } & $\Wjets$   &  $\Pe$       & \multirow{5}{*}{2 or 3}  & \multirow{5}{*}{2} &  & $50 < \mt < 160\GeV$\\
         \cline{1-1}\cline{4-4}
         bbG & & \semileptonic $\ttbar$   & $\mu$      &                              &                    &  & No selection on $\mindphi$ \\
        \cline{1-1}\cline{3-4}\cline{8-8}
                                                     bbH & & \multirow{2}{*}{$\Zjets$}               & $\ee$      &                              &                    &                                      & $70 < m_{\ell\ell} < 110\GeV$\\
        \cline{1-1}\cline{4-4}
                                                     bbI & &                                         & $\mumu$  &                              &                    &                                      & No selection on $\mindphi$ \\
        \cline{1-1}\cline{3-4}\cline{8-8}            bbJ & & Dileptonic $\ttbar$                                & $\emu$    &                              &                    &                                      & No selection on $\mindphi$ \\
        \hline
        \end{tabular}
        }
\end{table*}

\subsection{Signal region selections}\label{sec:selection_SR}
\textbf{Dileptonic \boldmath$\ttMET$:} Events in the dileptonic $\ttbar$ SR are required to contain exactly two leptons that satisfy stringent identification and isolation requirements.  One of the leptons must have $\pt > 30\GeV$, while the second must have $\pt > 10\GeV$.  Events containing additional, loosely identified leptons with $\pt > 10\GeV$ are rejected.  Events are also required to have $\ptmiss > 50\GeV$, and to contain two or more jets, at least one of which must satisfy b tagging requirements.  Overlaps between the dileptonic SR and the dileptonic and $\Zjets$ CRs of the \semileptonic $\ttMET$ and $\bbMET$ channels (discussed in Section~\ref{sec:selection_CR}) are removed by vetoing events that satisfy the selections for those CRs.  These vetoes remove 2.5\% of the events from the dileptonic $\ttMET$ SR.  The azimuthal opening angle between the $\pt$ vector of the dilepton system and the $\ptmiss$ vector, $\dphill$, is required to be larger than 1.2 radians.  This requirement preferentially selects events consistent with a $\ttbar$ system recoiling against the invisibly decaying DM mediator.  The dilepton mass, $\mll$, is required to be larger than $20\GeV$.  In dielectron and dimuon events, $\mll$ is also required to be at least $15\GeV$ away from the Z boson mass~\cite{pdg2016}.  These requirements reduce backgrounds from low-mass dilepton resonances and from leptonic Z boson decays.

Events that satisfy these criteria are divided among three SR categories that correspond to the flavor assignments of the two selected leptons: $\ee$, $\emu$, and $\mumu$.  Signal efficiencies for the dileptonic $\ttMET$ SR event selections range from $6\times 10^{-3}$ to $10^{-2}$ for mediator masses between $10\GeV$ and $500\GeV$.  The denominator used in the efficiency calculation is the total number of signal events, irrespective of the $\ttbar$ final state.  The low efficiencies result primarily from the small dileptonic branching fraction.

\textbf{\boldmath$\ell+\text{ jets } \ttMET$:} Events in the \semileptonic $\ttbar$ SR are selected by requiring $\ptmiss >160\GeV$, exactly one lepton, and three or more jets, of which at least one must satisfy the b tagging criteria. The lepton is required to have $\pt > 30\GeV$, and to pass tight identification criteria.  Events must not contain additional leptons with $\pt > 10\GeV$ that satisfy a looser set of identification requirements. To reduce SM \semileptonic $\ttbar$ and $\Wjets$ backgrounds, the transverse mass, calculated from $\ptvecmiss$ and the lepton momentum ($\vec{p}_{\mathrm{T}}^{\ell}$) as:
\begin{equation}
\mt=\sqrt{2\pt^{\ell}\ptmiss(1-\cos\Delta\phi(\vec{p}_{\textrm{T}}^{\ell},\ptvecmiss))},
\end{equation}
is required to be larger than $160\GeV$.

Following these selections, the remaining background events primarily consist of dileptonic $\ttbar$ final states in which one of the leptons is not identified.  Because of the requirement of $\ptmiss > 160\GeV$, this background tends to contain events with Lorentz-boosted top quark decays in which the b jet is closely aligned with the direction of the neutrino.  This background is suppressed by requiring that the smallest azimuthal angle formed from the missing transverse momentum vector and each of the two highest $\pt$ jets in the event, $\mindphi$ where $i=1,2$, be larger than 1.2 radians.  In addition, the $\mwt$ variable~\cite{MT2W} is required to be larger than $200\GeV$.  This variable is defined as:
\ifthenelse{\boolean{cms@external}}{
\begin{multline}
M_{\mathrm{T}2}^{\PW}  =  \min \Biggl\{ m_{y} \text{ consistent with: }\\
\Biggl[
\begin{aligned}
&\vec{p}_{1}^{T} + \vec{p}_{2}^{T} = \ptvecmiss, p_{1}^{2}=0, (p_{1} + p_{l})^{2} = p_{2}^{2} = M^{2}_{\PW}, \\
&(p_{1}+p_{l}+p_{b1})^{2} = (p_{2} + p_{b2})^{2} = m^{2}_{y}
\end{aligned}
\Biggr] \Biggr\}
\end{multline}
}{
\begin{equation}
M_{\mathrm{T}2}^{\PW}  =  \min \left\{ m_{y} \text{ consistent with: } \left[
\begin{aligned}
&\vec{p}_{1}^{T} + \vec{p}_{2}^{T} = \ptvecmiss, p_{1}^{2}=0, (p_{1} + p_{l})^{2} = p_{2}^{2} = M^{2}_{\PW}, \\
&(p_{1}+p_{l}+p_{b1})^{2} = (p_{2} + p_{b2})^{2} = m^{2}_{y}
\end{aligned}
\right] \right\}
\end{equation}
}
where $m_{y}$ is the mass of two parent particles that each decay to bW($\ell\nu$).  One of the W decays is assumed to produce a lepton that is not reconstructed.  For the W decay that does produce a reconstructed lepton, the neutrino and lepton 4-momenta are denoted $p_{1}$ and $p_{\ell}$, respectively.  The 4-momentum of the W that produces the unreconstructed lepton is denoted $p_{2}$, while the momenta of the two b candidates are referred to as $p_{b1}$ and $p_{b2}$.  Assuming perfect measurements, the $\mwt$ has a kinematic end-point at \mtop~ for $\ttbar$ events, whereas signal events lack this feature because both the neutrino and DM particles contribute to $\ptmiss$.

The efficiency of the \semileptonic $\ttMET$ SR event selections for the $\ttDM$ process range from $10^{-4}$ for mediator masses of the order of $10\GeV$, to $10^{-3}$ for masses of about 500\GeV.  Signal efficiencies are low because of the stringent $\ptmiss$ requirement applied.  The efficiency improves with increasing mediator mass because of the broadening of the $\ptmiss$ spectrum.

\textbf{All-hadronic \boldmath$\ttMET$:} Any event with a loosely identified lepton with $\pt > 10\GeV$ is vetoed from the all-hadronic $\ttMET$ SRs. The $\ptmiss$ value must be larger than $200\GeV$, and four or more jets are required, at least one of which must satisfy b tagging criteria.  Spurious $\ptmiss$ can arise in multijet events due to jet energy mismeasurement.  In such cases, the reconstructed $\ptmiss$ tends to align with one of the jets.  Multijet background is suppressed by requiring that $\mindphi>0.4$  or 1 radian (depending on the number of RTT tags, as described below) for all jets in the event.  The $\mindphi$ selections also help to reduce \semileptonic $\ttbar$ background, for which the $\ptmiss$ vector is typically aligned with a b jet.

Following these selection requirements, the dominant residual background is \semileptonic SM $\ttbar$ production.  By contrast, selected signal typically includes events in which both top quarks decay hadronically.  The resolved top quark tagger (RTT, introduced in Section~\ref{sec:detector}) is employed to suppress the \semileptonic background by identifying potential hadronic top quark decays.  The RTT is applied to the all-hadronic search region to define a category of events with two hadronic top quark decays.  In this double-tag (2RTT) category, one or more b-tagged jets are required and $\mindphi>0.4$ radians is imposed for all jets in the event.  The 2RTT category implicitly requires at least six jets in the event.  A second category is defined for events with 0 or 1 top quark tags (0,1RTT), four or more jets with at least two b-tagged jets, and a tighter requirement of $\mindphi>1$ radian.

The selection efficiency for $\ttDM$ events in the all-hadronic $\ttMET$ SRs ranges from $10^{-3}$ for mediator masses of the order of 10\GeV to $10^{-1}$ for masses near $500\GeV$.  These values are larger than the corresponding efficiencies of the dileptonic and \semileptonic SR selections because of the larger branching fraction to the all-hadronic final state.

{\tolerance=2400
\textbf{{\boldmath$\bbMET$}:} Events with $\ptmiss>200\GeV$ are selected for the SRs of this final state.  Events containing identified and isolated electrons or muons with $\pt$ larger than 10\GeV or identified $\tau$ leptons with $\pt > 18\GeV$ are rejected.  Multijet background is reduced by requiring $\mindphi > 0.5$ radians for all jets in the event.
\par}

Following these selections, two exclusive event categories are defined using the number of jets and b-tagged jets in the event.  The single b-tagged jet category provides high efficiency for $\bbDM$ signal and requires at most two jets.  At least one of these jets must have $\pt > 50\GeV$, and exactly one must satisfy b tagging requirements.  The second category allows exactly two b-tagged jets.  This SR selects $\bbDM$ signal and partially recovers $\ttDM$ events that are not selected in the all-hadronic $\ttMET$ categories.  At most three jets are allowed in the 2 b tag SR, and at least two of these jets must have $\pt > 50\GeV$.

The efficiency of the $\bbMET$ SR event selections for the $\bbDM$ process range from $10^{-6}$ for mediator masses of the order of 10\GeV, to $10^{-2}$ for masses of 500\GeV.  The selection efficiency for the $\ttDM$ process is found to be less dependent on the mediator mass, and varies from $10^{-4}$ to $10^{-3}$ for the same mass range.

\subsection{Background control region selections}\label{sec:selection_CR}
\begin{figure}[hbt]
\centering
\includegraphics[width=0.49\textwidth]{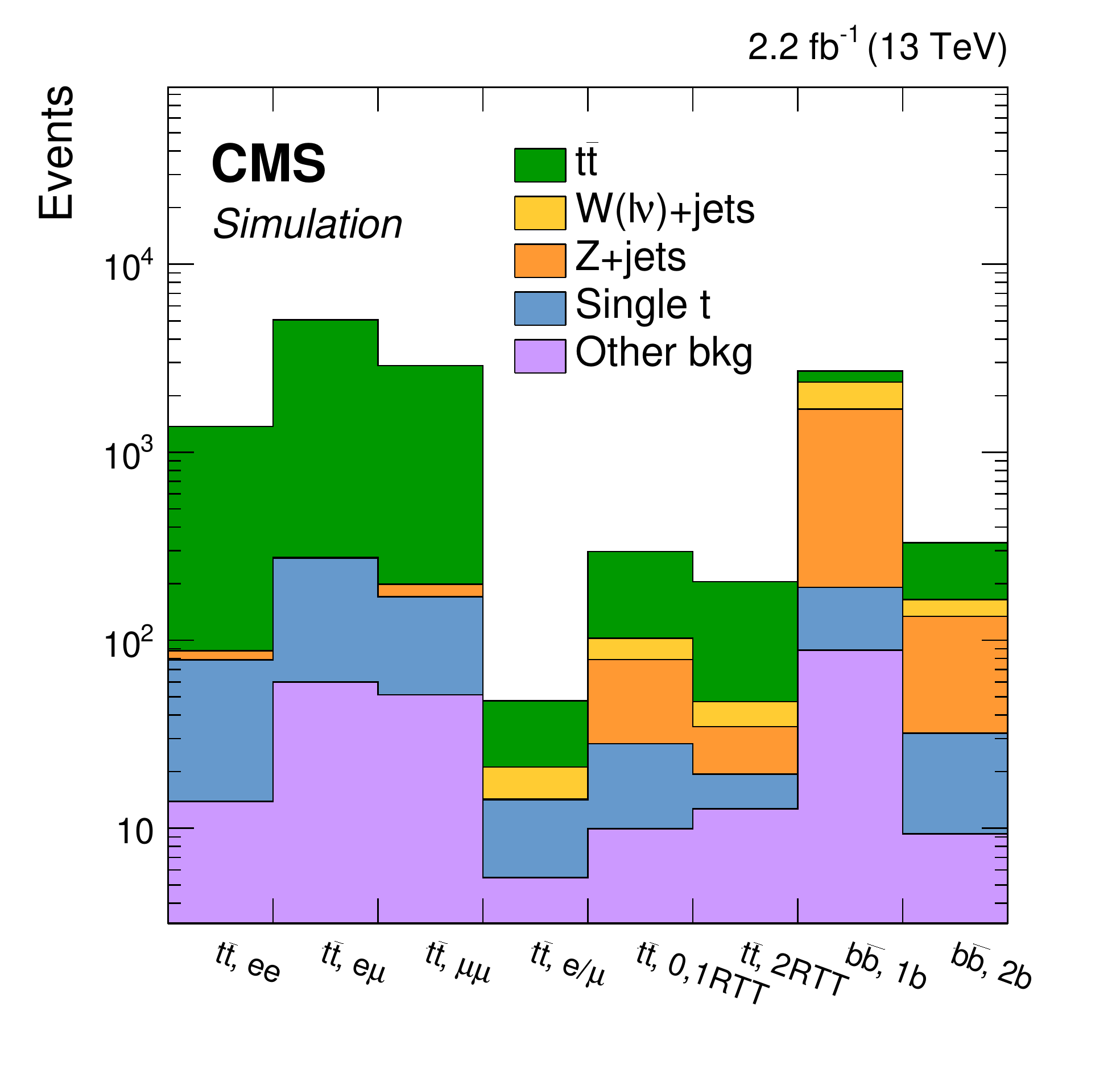}
\caption{Simulation-derived background expectations in the $\ttMET$ and $\bbMET$ signal regions.}
\label{fig:SR}
\end{figure}

Figure~\ref{fig:SR} shows the simulated background yields in each of the SRs following the selections of Section~\ref{sec:selection_SR}.  Clearly, the dominant backgrounds in the SRs are from the SM $\ttbar$, $\Wjets$, and $\Zjets$ processes.  The estimation of backgrounds in the SRs is improved through the use of corresponding data CRs enriched in these processes.  Independent CRs are defined for each of the \semileptonic $\ttMET$, all-hadronic $\ttMET$ and $\bbMET$ SRs.  In some cases, multiple CRs are used to constrain a given background process in a SR.  In this section we describe the main $\ttbar$, $\Wjets$, and $\Zjets$ backgrounds and the selections used to define the CRs.  The CR selections are designed to ensure that these regions are both mutually exclusive and exclusive of the SRs as well.  The contributions of multijet, diboson, single t, and $\ttbar+{\Z/\PW/}\gamma$ processes in the SRs are either subdominant or insignificant after the SR selections.  The residual backgrounds from these processes are modeled with simulation.  Dilepton background events from Drell--Yan and processes in which jets are misidentified as leptons are estimated using the sideband techniques described in Ref.~\cite{Chatrchyan:2012ty}.

The remainder of this section describes how the contributions of SM backgrounds in the SRs are estimated using the CRs.  The discussion utilizes the CR labeling convention defined in Table~\ref{tab:selections_CR}, for ease of reference.  The CRs for the \semileptonic $\ttMET$ SR are denoted slA and slB, those for the all-hadronic $\ttMET$ SRs are hadA--hadG, and those for the $\bbMET$ SRs are bbA--bbJ.

\begin{figure}[!htb]
\centering
  \includegraphics[width=0.49\textwidth]{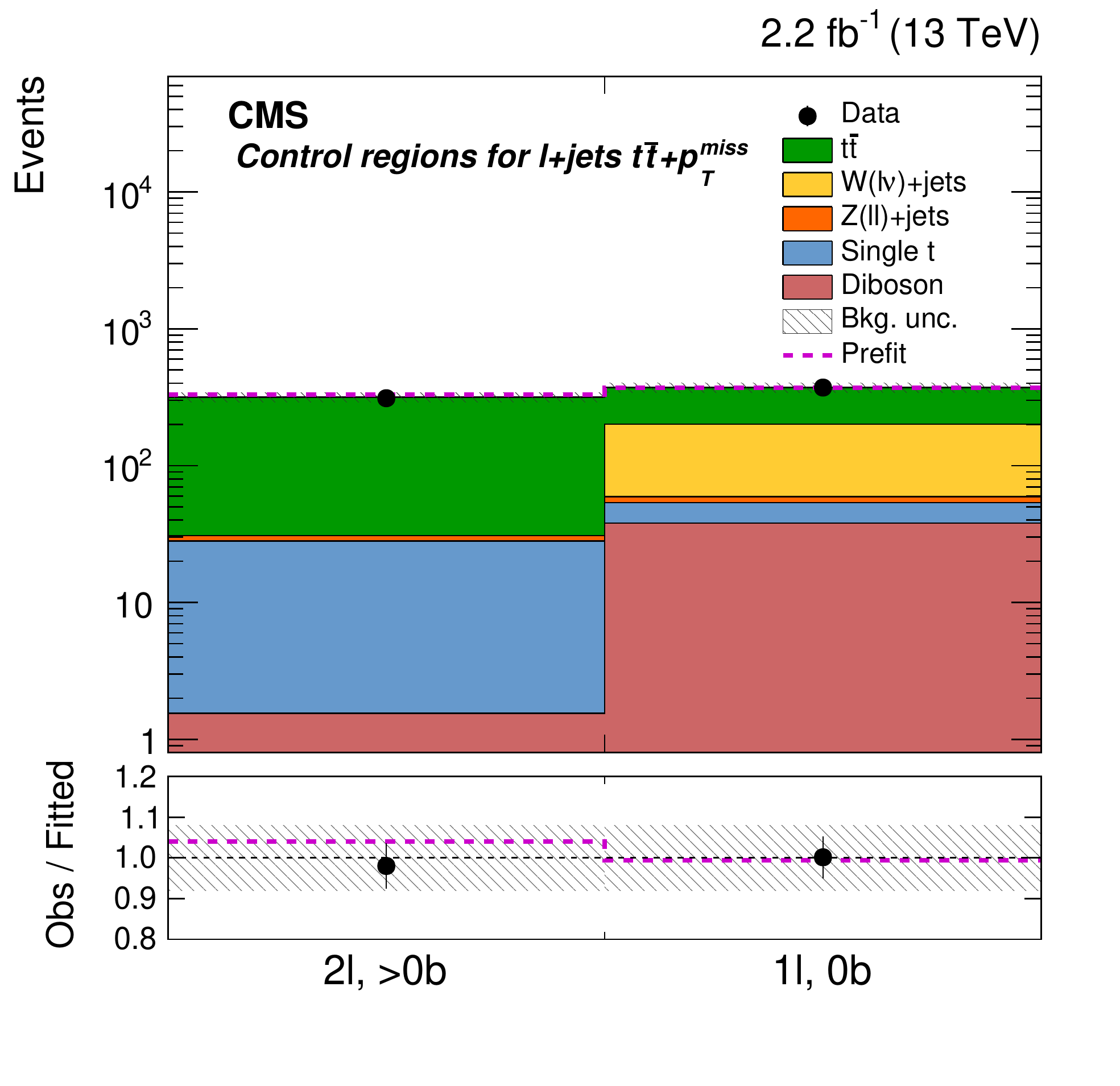}
  \caption{Observed data, and prefit and fitted background-only event yields in the control regions associated with the \semileptonic $\ttMET$ signal region.  The 2 lepton, $\geq$0 b tag region (slA in Table~\ref{tab:selections_CR}) is used to constrain the dileptonic $\ttbar$ background in the \semileptonic $\ttMET$ signal region, while the 1 lepton, 0 b tag control region (slB) constrains $\Wjets$ background.  The lower panel shows the ratios of observed to fitted background yields.  In both panels, the statistical uncertainties of the data are indicated as vertical error bars and the fit uncertainties as hatched bands.  Prefit yields and the ratios of prefit to fitted background expectations are shown as dashed magenta histograms.}
  \label{fig:prefit_yields_DLSL}
\end{figure}

\begin{figure}[!hbtp]
\centering
  \includegraphics[width=0.49\textwidth]{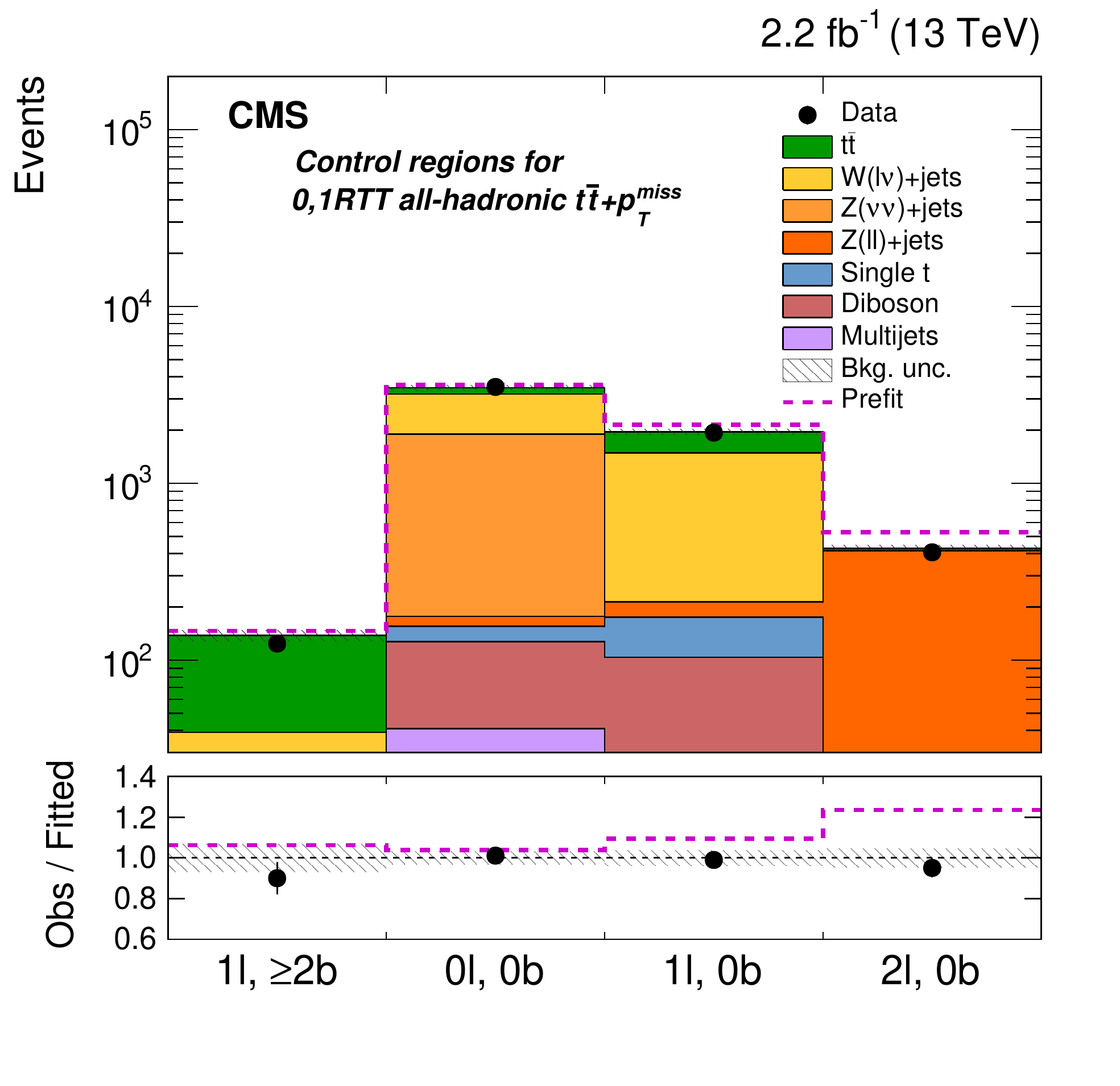}
  \includegraphics[width=0.49\textwidth]{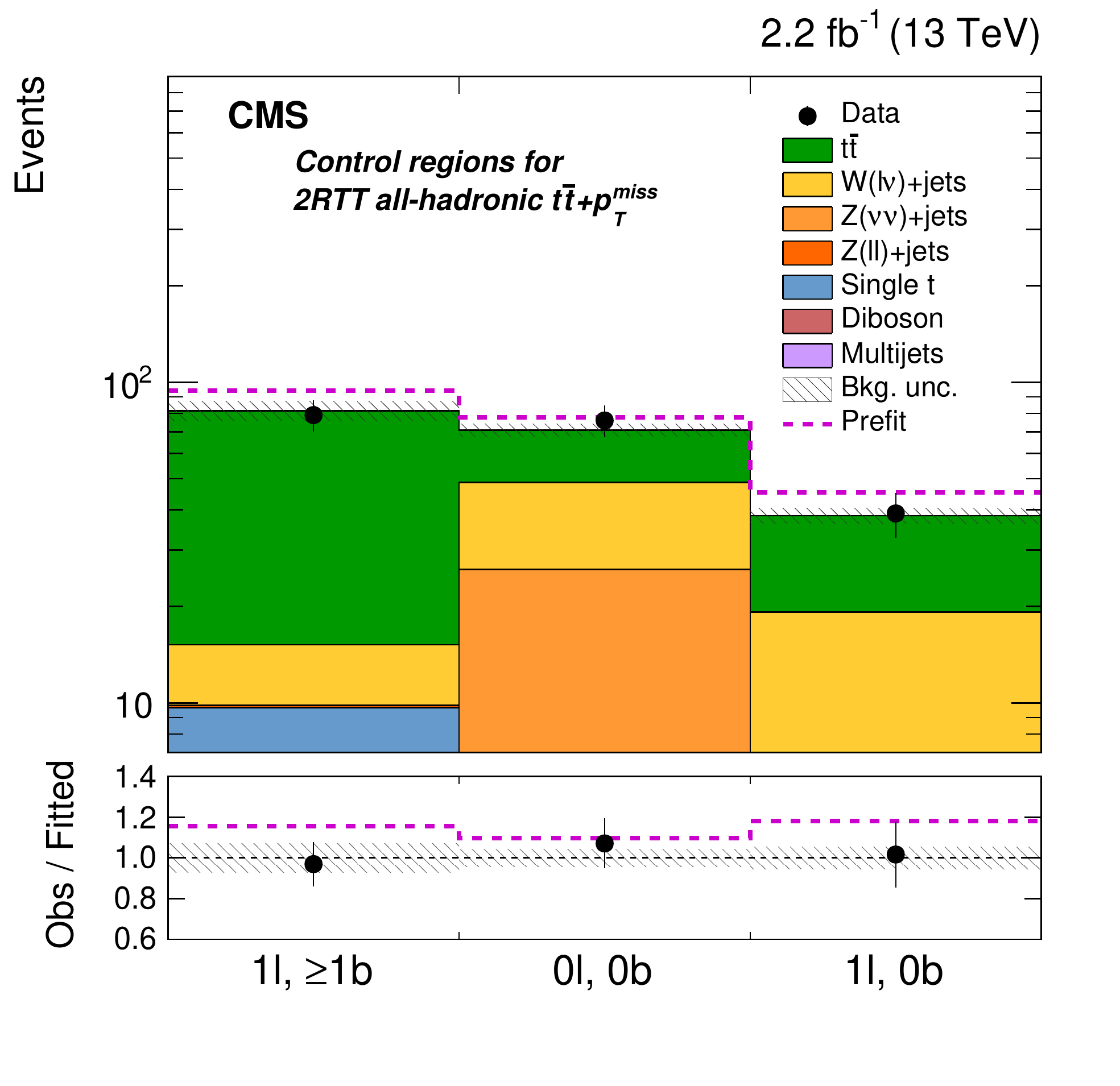}
  \caption{Observed data, and prefit and fitted background-only event yields in the control regions associated with the 0,1RTT (\cmsLeft) and 2RTT (\cmsRight) all-hadronic $\ttMET$ signal regions.  The 1 lepton, ${\geq}2$ b tag control region (hadA in Table~\ref{tab:selections_CR}) constrains \semileptonic $\ttbar$ background in the 0,1RTT signal region.  This process is constrained in the 2RTT signal region using the 1 lepton, $\geq$1 b tag control region (hadE).  The $\le$1 lepton, 0 b tag control regions (hadB, hadC, hadF, hadG) constrain $\Wjets$ and $\Zjets$ backgrounds, while the 2 lepton, 0 b tag control region (hadD) provides an additional constraint on the $\Zjets$ background.  The lower panels show the ratios of observed to fitted background yields.  In both panels, the statistical uncertainties of the data are indicated as vertical error bars and the fit uncertainties as hatched bands.  Prefit yields and the ratios of prefit to fitted background expectations are shown as dashed magenta histograms.}
  \label{fig:prefit_yields_tthad}
\end{figure}

\begin{figure}[!hbt]
\centering
  \includegraphics[width=0.49\textwidth]{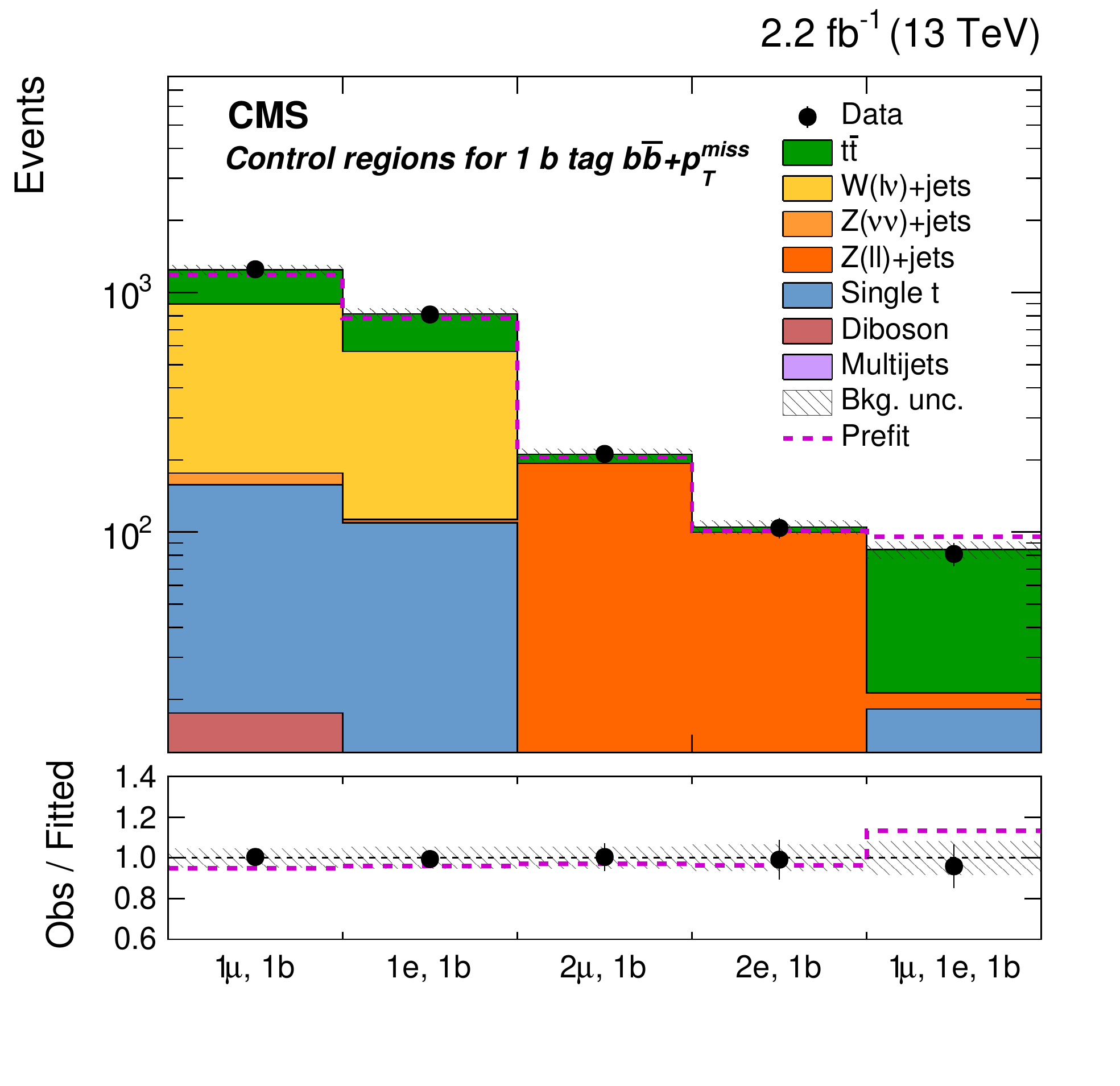}
  \includegraphics[width=0.49\textwidth]{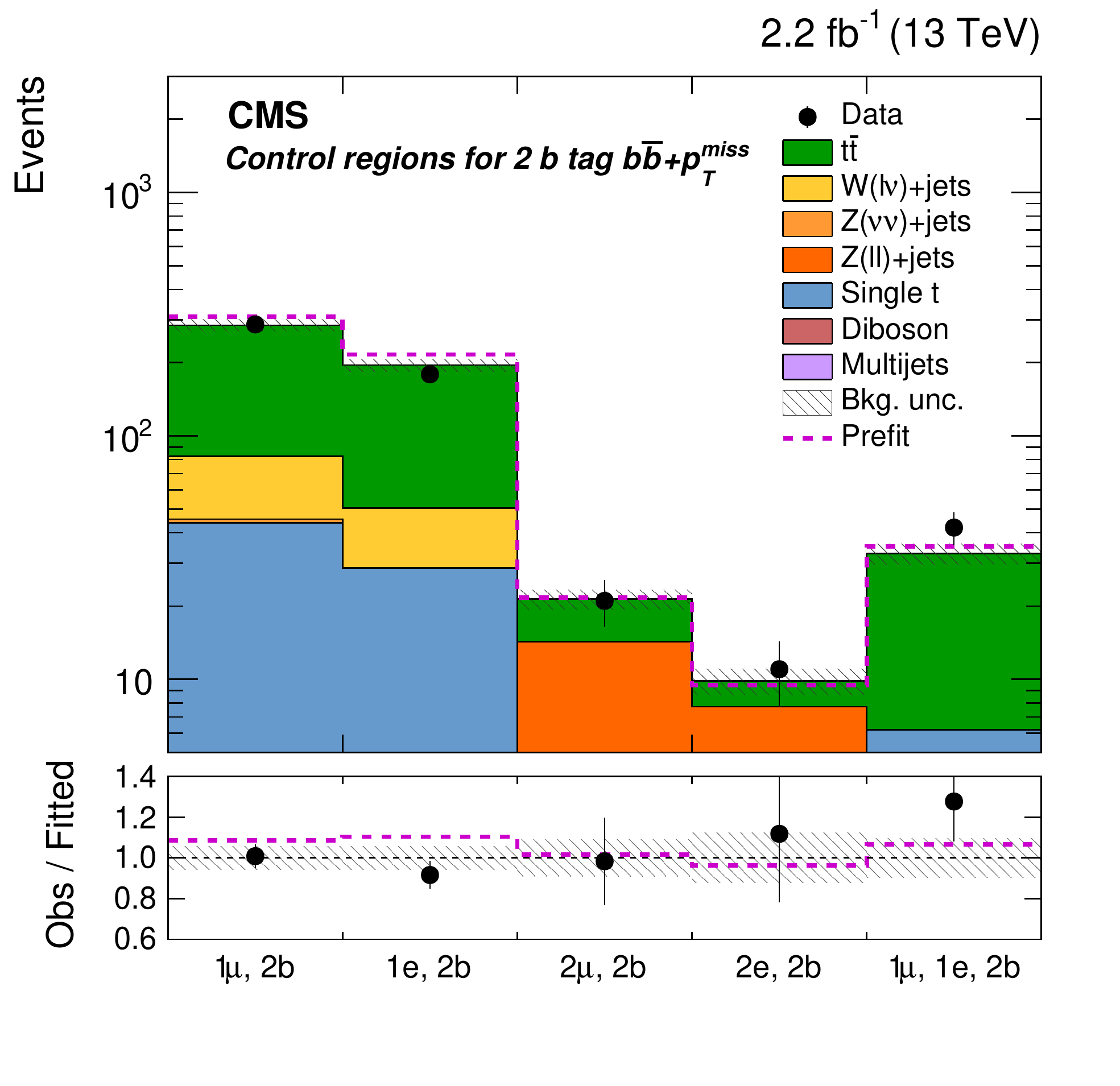}
   \caption{Observed data, and prefit and fitted background-only event yields in the control regions associated with the $\bbMET$ signal region with 1 b tag (\cmsLeft) and with 2 b tags (\cmsRight).  The 1 lepton, $\geq$1 b control regions (bbA, bbB, bbF and bbG in Table~\ref{tab:selections_CR}) are used to constrain $\Wjets$ and $\ttbar$ backgrounds in the $\bbMET$ signal regions.  The dileptonic control regions (bbC-bbE, bbH-bbJ) are used to constrain $\Zjets$ and $\ttbar$ backgrounds.  The lower panels show the ratio of observed to fitted background yields.  In both panels, the statistical uncertainties of the data are indicated as vertical error bars and the fit uncertainties as hatched bands.  Prefit yields and the ratios of prefit to fitted background expectations are shown as dashed magenta histograms.}
  \label{fig:prefit_yields_bb}
\end{figure}

Section~\ref{sec:extract} describes how the CRs are simultaneously fit with the SRs to constrain the predicted normalization of the $\ttbar$, $\Wjets$, and $\Zjets$ background processes.  Figures~\ref{fig:prefit_yields_DLSL}--\ref{fig:prefit_yields_bb} compare the integrated yields in each CR before and after background-only fits to the CR $\ptmiss$ distributions.  Reasonable agreement is found between the observed and predicted CR yields.  In general, the expected and observed $\ptmiss$ distributions in the CRs also agree.  Regions for which the distributions of data and of the initial (``prefit'') MC disagree are noted in the text.

\textbf{Dileptonic \boldmath$\ttbar$:} Dileptonic $\ttbar$ background in the \semileptonic $\ttbar$ SR consists of events in which only one of the leptons is identified.  A dileptonic CR (slA) for the \semileptonic $\ttMET$ search region is defined by requiring an additional lepton with respect to the \semileptonic selection, and by removing the selections on $\mt$, $\mwt$, and $\mindphi$.  Both leptons from dileptonic $\ttbar$ decays in the \semileptonic SR are typically within the detector acceptance.  The lepton momenta are therefore included in the $\pt$ vector sum for this CR, so as to simulate the $\ptmiss$ distribution expected for the dileptonic $\ttbar$  background in the \semileptonic SR.   Mutual exclusion with the dileptonic $\ttbar$ and $\Zjets$ CRs of the $\bbMET$ search region (described below) is ensured by vetoing events that additionally satisfy the selection requirements of those CRs.

The $\ttbar$ background in the $\bbMET$ SRs consists of dileptonic and \semileptonic $\ttbar$ events in which no leptons are identified.  Dileptonic $\ttbar$ CRs (bbE, bbJ) are formed for the 1 b tag and 2 b tag $\bbMET$ SRs by requiring two opposite-charge, different-flavor leptons with $\pt>30\GeV$.  Tight (loose) identification and isolation criteria are imposed on the leading $\pt$ (subleading $\pt$) lepton.  In contrast to the dileptonic background in the \semileptonic $\ttMET$ SR, the leptons from $\ttbar$ in the $\bbMET$ SRs typically fall outside of the detector acceptance.  The momentum of the selected leptons in the $\bbMET$ CRs is therefore subtracted from the $\ptmiss$ observable in order to mimic the $\ptmiss$ distribution in the SR.  The SR requirements on $\mindphi$, which primarily remove multijet background, are not imposed.  All other selections from the $\bbMET$ SRs are applied.

Dileptonic $\ttbar$ production is the dominant SM background in the dileptonic $\ttMET$ SRs.  Corresponding CRs are not employed for this search channel because dileptonic $\ttbar$ events are found to be well-modeled by simulation and are selected with high efficiency in the dileptonic SR.

\textbf{\boldmath $\ell+\text{ jets } \ttbar$:} The most significant source of background in the hadronic $\ttMET$ SRs is \semileptonic $\ttbar$ production.  This process contributes to the hadronic $\ttMET$ search when the lepton is not identified. Control regions for \semileptonic $\ttbar$ (hadA, hadE) are defined by selecting events with exactly one identified lepton with $\pt>30\GeV$, and by requiring $\mt<160\GeV$ in order to avoid overlaps with the SR of the \semileptonic channel.  All other requirements used to define the hadronic SRs are applied, and the CR is split into 0,1RTT and 2RTT categories.

The dileptonic $\ttbar$ CRs for the $\bbMET$ search (described above) provide stringent constraints on $\ttbar$ backgrounds in the corresponding SRs.  Additional constraints on $\ttbar$ background in this channel are provided through four single-lepton CRs (bbA, bbB, bbF, and bbG).  A single-electron (muon) CR for the 1 b tag SR requires exactly one electron (muon) with $\pt>30\GeV$.  The lepton must satisfy tight isolation and identification criteria.  The $\mt$ observable calculated from the lepton momenta and $\ptmiss$ must satisfy $50 < \mt < 160\GeV$.  Except for the requirement on $\mindphi$, each of the selection criteria for the 1 b tag signal category must also be satisfied.  Analogous CRs for the 2 b tag signal category are formed by applying the corresponding signal selection criteria.  As in the dileptonic $\ttbar$ CRs for the $\bbMET$ searches, the lepton is removed from the $\ptmiss$ calculation.

{\tolerance=1800
\textbf{\boldmath $\Wjets$:} A \Wjets~CR for the \semileptonic $\ttMET$ search (slB) is created by requiring zero b tags. The $\mt>160\GeV$ requirement from the \semileptonic signal selection is maintained, however, the cuts on $\mwt$ and $\mindphi$ are removed.
\par}

\begin{figure}[!hbt]
\centering
  \includegraphics[width=0.49\textwidth]{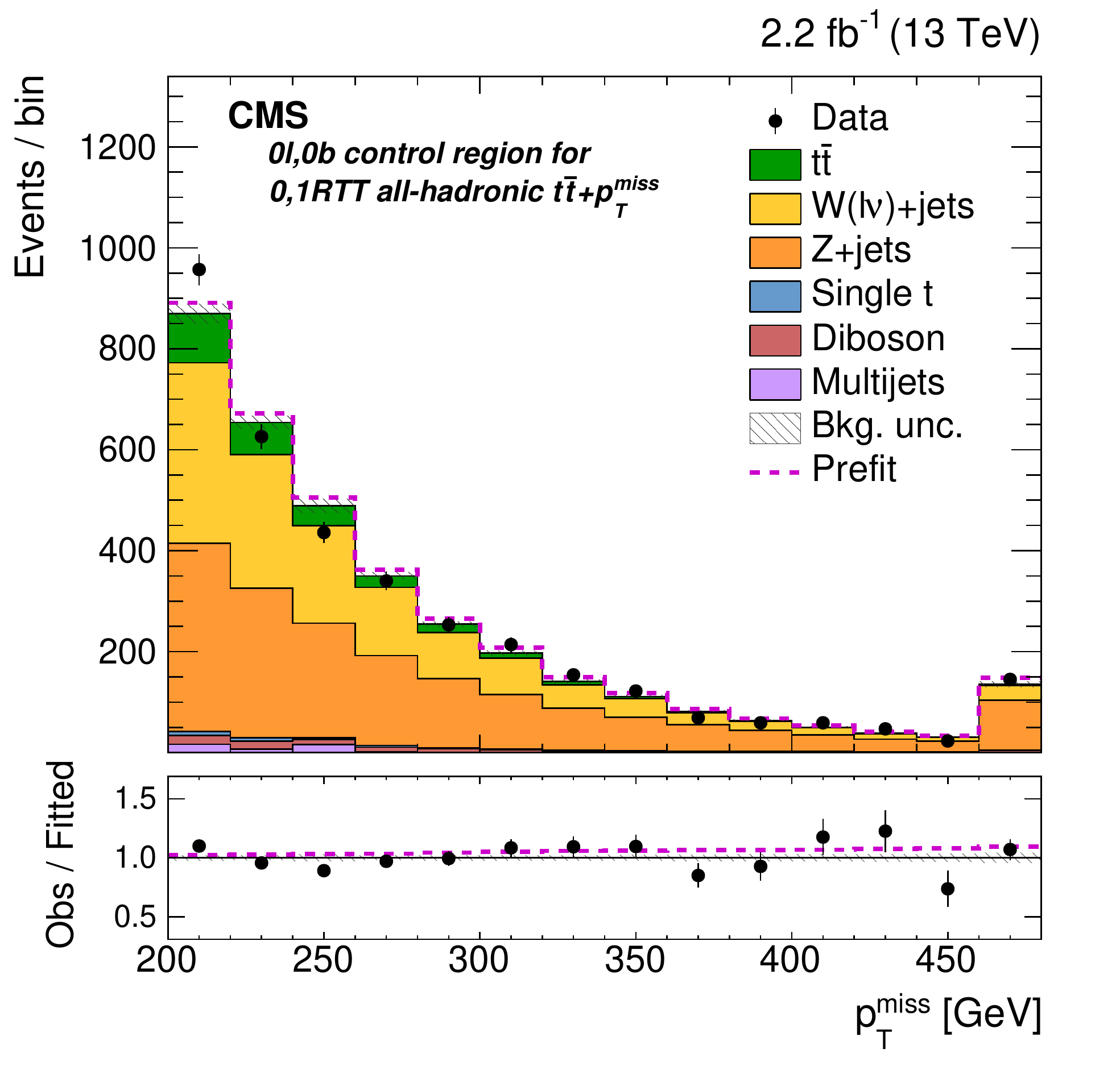}
  \includegraphics[width=0.49\textwidth]{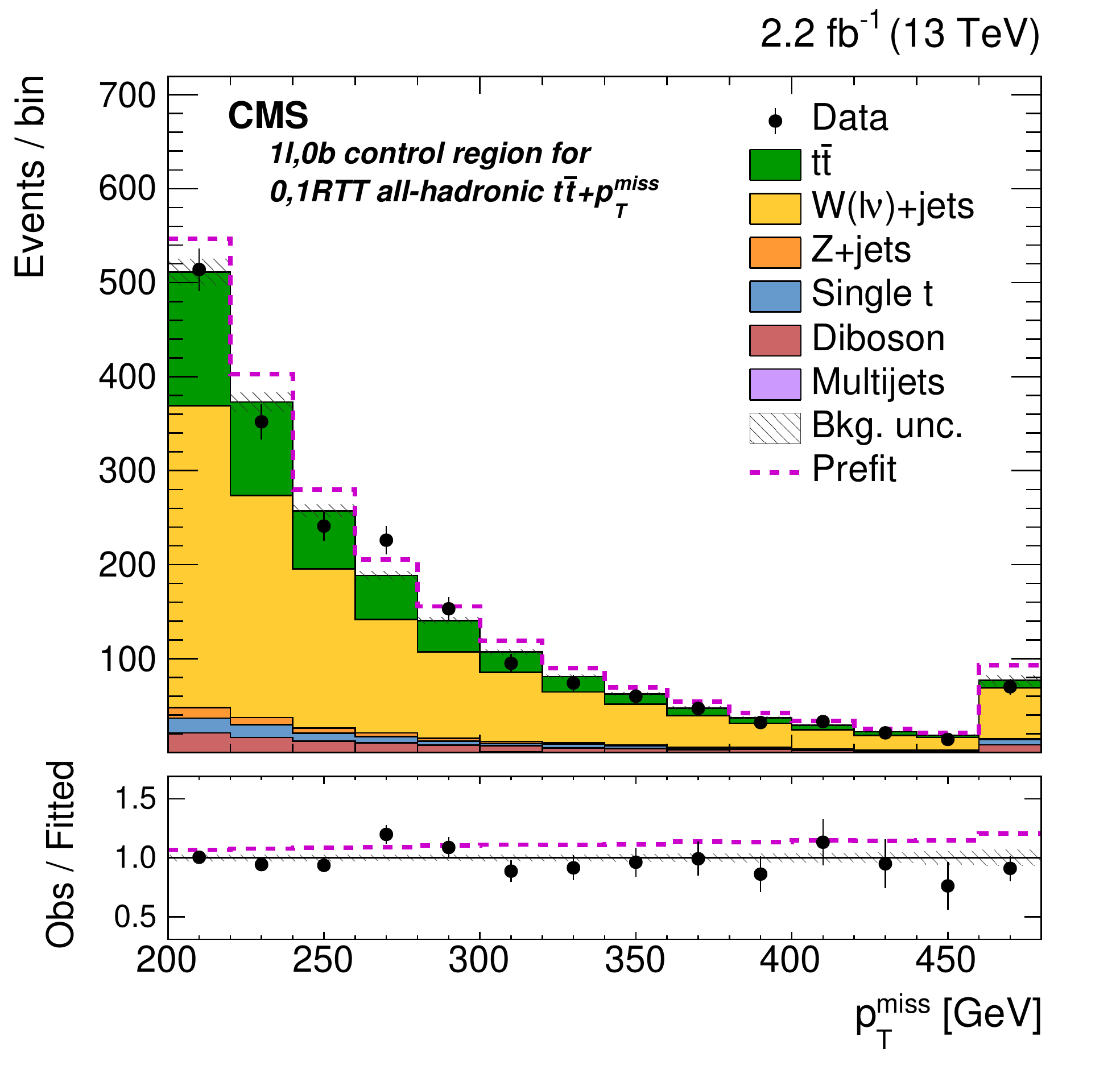}
  \caption{Observed data, and prefit and fitted background-only \ptmiss~distributions in two control regions (hadB and hadC in Table~\ref{tab:selections_CR}) for the 0,1RTT hadronic $\ttMET$ signal region with 0 leptons (\cmsLeft) and with 1 lepton (\cmsRight) and 0 b tags.  The 0 lepton control region is used to constrain $\Wjets$ and $\Zjets$ backgrounds.  The 1 lepton CR provides an additional constraint on $\Wjets$ background.  The last bin contains overflow events.  The lower panels show the ratios of observed data to fitted background yields.  In both panels, the statistical uncertainties of the data are indicated as vertical error bars and the fit uncertainties are indicated as hatched bands.  Prefit yields and the ratios of prefit to fitted background expectations are shown as dashed magenta histograms.}
  \label{fig:prefit_ptmiss_vwjetsCR_01RTT}
\end{figure}

Control regions enriched in both $\Wjets$ and $\Zjets$ (hadB, hadF) are formed for the all-hadronic $\ttMET$ categories by modifying the SR selections to require zero b tags.  In addition, dedicated $\Wjets$ CRs (hadC, hadG) are defined by requiring the presence of an isolated, identified lepton with $\pt > 30\GeV$ and $\mt < 160\GeV$.  The \WZjets~and \Wjets~CRs are both categorized using the number of RTTs, as in the corresponding SRs.  The prefit yields and $\ptmiss$ distributions in the hadB and hadC regions are observed to differ from those of data.  The discrepancy is due to a mismodeling of hadronic activity in the simulation, which leads to an overestimation of the selection efficiency for the Z+jets and W+jets processes.  Reasonable agreement is achieved through the fit, as is shown in Figs~\ref{fig:prefit_ptmiss_vwjetsCR_01RTT}. and~\ref{fig:prefit_yields_tthad}.

The $\Wjets$ process contributes the second-largest background in the 1 b tag SR of the $\bbMET$ channel.  This background is constrained via the single-lepton CRs (bbA, bbB, bbF, bbG) of the $\bbMET$ channel, which were introduced previously in the context of constraints on \semileptonic $\ttbar$ backgrounds.

\begin{figure}[!hbt]
\centering
  \includegraphics[width=0.49\textwidth]{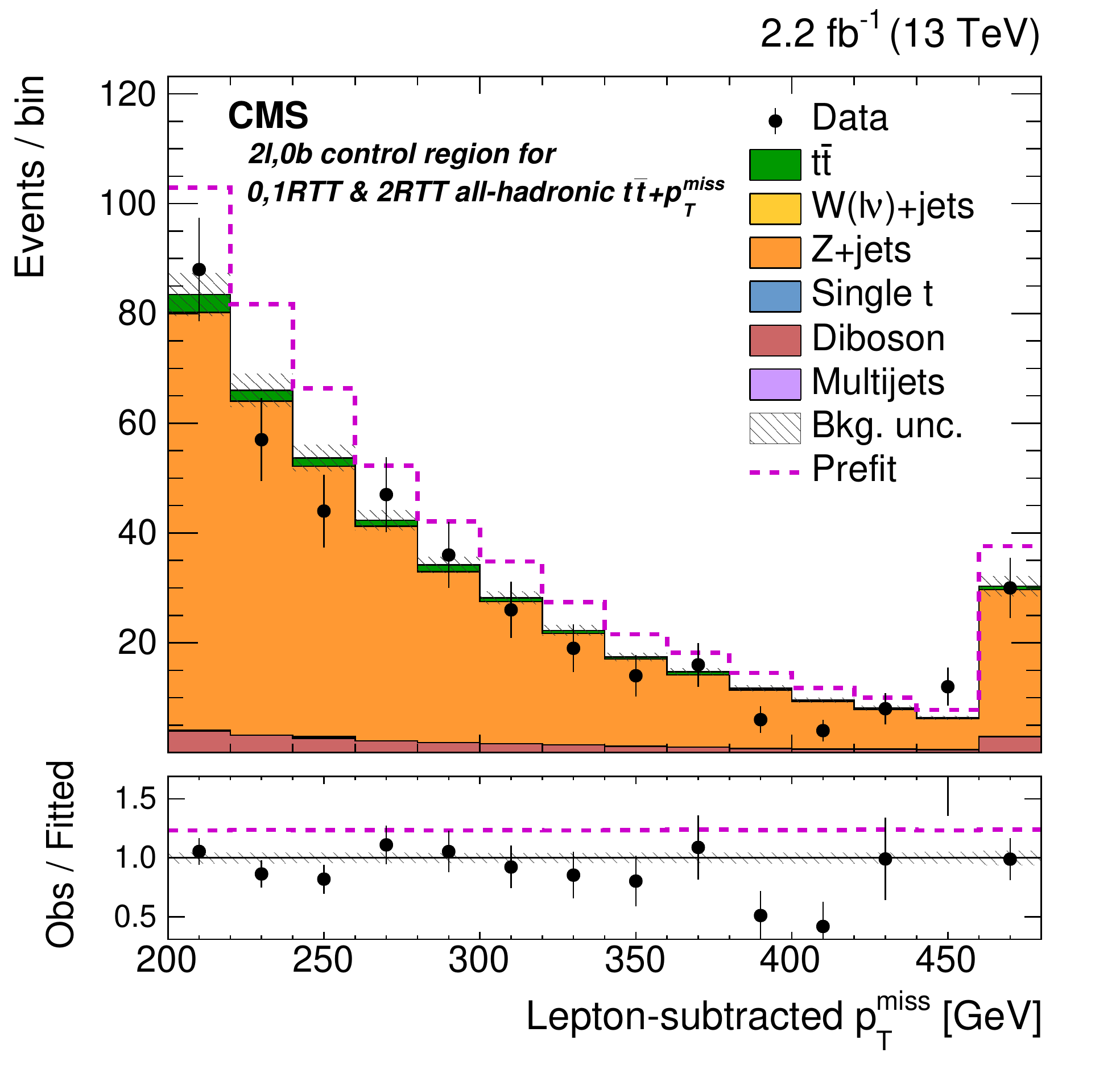}
  \caption{Observed data, and prefit and fitted background-only, lepton-subtracted \ptmiss~distributions in the dileptonic control region (hadD in Table~\ref{tab:selections_CR}) for the all-hadronic $\ttMET$ signal regions.  This control region is used to constrain $\Znunu$ background.  The selections for jets and $\ptmiss$ used in the 0,1RTT signal region are applied, with those on $\ptmiss$ applied to lepton-subtracted $\ptmiss$.  The signal region requirements on $\mindphi$ and b tags are removed to increase $\Zjets$ yields.  The last bin contains overflow events.  The lower panel shows the ratios of observed data to fitted background yields.  In both panels, the statistical uncertainties of the data are indicated as vertical error bars and the fit uncertainties are indicated as hatched bands.  Prefit yields and the ratios of prefit to fitted background expectations are shown as dashed magenta histograms.}
  \label{fig:prefit_ptmiss_zllCR_hadronic}
\end{figure}

\textbf{\boldmath$\Zjets$:} The $\Znunu$ process is a significant source of background in the all-hadronic $\ttMET$ SRs.  This background is partially controlled via the $\WZjets$ CRs (hadB, hadF) described previously.  An additional constraint is derived from a distinct $\Zll$ CR (hadD), in which two oppositely-charged, same-flavor leptons are required to pass tight isolation and identification requirements.  The mass of the lepton pair must fall between 60 and 120\GeV.  A prediction for the $\ptmiss$ distribution in the hadronic SRs is obtained by subtracting the lepton momenta in the $\ptmiss$ calculation.  The $\Zll$ CR is not categorized in the number of RTTs because of the negligible yields obtained with two RTT tags.  The selections for jets and $\ptmiss$ used in the 0,1RTT SR are applied in the $\Zll$ CR, with those on $\ptmiss$ applied to lepton-subtracted $\ptmiss$. The requirements on $\mindphi$ and b tags are removed to increase $\Zjets$ yields.   Figure~\ref{fig:prefit_ptmiss_zllCR_hadronic} demonstrates that the lepton-subtracted $\ptmiss$ distribution observed in the $\Zll$ CR of the all-hadronic channel is not well described by the prefit expectation.  Agreement substantially improves following the fit.

The $\Znunu$ process is also a significant background in the $\bbMET$ SRs.  This background is constrained with four distinct CRs: bbC, bbD, bbH, and bbI.  The $\Zee$ and $\Zmm$ CRs require two electrons and two muons with $\pt>30\GeV$, respectively.  The isolation and identification criteria applied on the leading-$\pt$ lepton are identical to those used in the \Wjets~CRs for the $\bbMET$ channel.  The subleading lepton is required to satisfy a looser set of isolation and identification criteria, as in the dileptonic CRs.  The leptons must be consistent with the decay of a Z boson; opposite-charge, same-flavor requirements are imposed, and the leptons must satisfy a constraint on the dilepton mass of $70 < \mll < 110\GeV$.  As in the \Wjets~and dileptonic $\ttbar$~CRs, events must also satisfy all but the $\mindphi$ selection criteria of the corresponding 1 b tag or 2 b tag signal category.  As in the $\Zjets$ CR for all-hadronic $\ttbar$ channel, lepton momenta are subtracted in the $\ptmiss$ calculation to approximate the distribution of $\ptmiss$ from \Znunu~expected in the $\bbMET$ SRs.

\section{Signal extraction}\label{sec:extract}
A potential DM signal could be revealed as an excess of events relative to SM expectations in a region of high $\ptmiss$.  The shape of the observed \ptmiss~distribution provides additional information that is used in this analysis to improve the sensitivity of the search.  A potential signal is searched for via simultaneous template fits to the \ptmiss~distributions in the SRs and the associated CRs defined in Sections~\ref{sec:selection_SR} and~\ref{sec:selection_CR}.  Signal and background $\ptmiss$ templates are derived from simulation and are parameterized to allow for constrained shape and normalization variations in the fits.

The fits are performed using the \textsc{RooStats} statistical software package~\cite{RooStats}.  The effects of uncertainties in the normalizations and in the $\ptmiss$ shapes of signal and background processes are represented as nuisance parameters.  Uncertainties that only affect normalization are modeled using nuisance parameters with log-normal probability densities.  Uncertainties that affect the shape of the $\ptmiss$ distribution, which may also include an overall normalization effect, are incorporated using a template ``morphing'' technique.  These treatments, as well as the approach used to account for MC statistical uncertainties on template predictions, follow the procedures described in Ref.~\cite{Conway-PhyStat}.

Within each search channel, additional unconstrained nuisance parameters scale the normalization of each dominant background process ($\ttbar$, $\Wjets$, and $\Zjets$) across the SRs and CRs.  For example, a single parameter is associated with the contribution of the \semileptonic $\ttbar$ process in the all-hadronic $\ttMET$ SRs and CRs.  A separate parameter is associated with the \semileptonic $\ttbar$ background in the $\bbMET$ SRs and CRs.  These nuisance parameters allow the data in the background-enriched CRs to constrain the background estimates in the SRs to which they correspond.  Because separate nuisance parameters are used for each search channel, a given normalization parameter cannot affect background predictions in unassociated search channels.  The yields and $\ptmiss$ shapes of subdominant backgrounds vary in the fit only through the constrained nuisance parameters.  Signal yields in the SRs and associated CRs are scaled simultaneously by signal strength parameters ($\mu$), defined as the ratio of the signal cross section to the theoretical cross section, $\mu=\sigma/\sigma_\text{TH}$.  The $\mu$ parameters scale signal normalization coherently across regions, and thus account for signal contamination in the CRs.

Signal extraction is performed for the individual search channels as well as for their combination.  The separate fits to the individual signal and associated CRs provide independent estimates of $\bbDM$ and $\ttDM$ contributions in each channel.   In this fitting scenario, separate signal strength parameters are used for each of the search channels.  The $\bbDM$ process is considered as a potential signal in the 1 b tag and 2 b tag regions of the $\bbMET$ channel.  The $\ttDM$ process is searched for in all SRs of the $\bbMET$ and $\ttMET$ channels separately.  The contribution of the $\bbDM$ process in the all-hadronic $\ttMET$ channel is negligible due to the jet multiplicity requirement.  An inclusive fit to all signal and CRs is also performed.  This fit uses a single signal strength parameter to extract the combined contribution of $\ttDM$ and $\bbDM$ in data.  Additional details on the per-channel and combined fits are provided in Section~\ref{sec:results}.

\section{Systematic uncertainties}\label{sec:syst}
Table~\ref{tab:syst} summarizes the uncertainties considered in the signal extraction fits.  The procedures used to evaluate the uncertainties are described later in this section.  Normalization uncertainties are expressed relative to the predicted central values of the corresponding nuisance parameters.  These uncertainties are used to specify the widths of the associated log-normal probability densities.  The integrated luminosity, b tagging efficiency, $\ptmiss$ trigger efficiency, pileup, and multijet/single t background normalization uncertainties are taken to be fully correlated across SRs and CRs.  Shape uncertainties are expressed in Table~\ref{tab:syst} as the change in the prefit yields of the lowest and highest $\ptmiss$ bins resulting from a variation of the corresponding nuisance by $\pm$1 standard deviation (s.d.).  These uncertainties are propagated to the fit by using the full $\ptmiss$ spectra obtained from $\pm$1 s.d. variations of the corresponding nuisance parameters~\cite{Conway-PhyStat}.  The PDF and jet energy scale shape uncertainties are taken to be fully correlated across SRs and CRs.  In general, the uncertainty estimation is performed in the same way for signal and background processes; however, the uncertainty from missing higher-order corrections for signal processes, which is approximately 30\% at LO in QCD, is not considered to facilitate a comparison with other CMS DM results.

\begin{table*}[phtb]
\centering
\topcaption{Summary of systematic uncertainties in the signal regions of each search channel.  The values given for uncertainties that are not process specific correspond to the dominant background in each signal region (i.e. $\Zjets$ in the 1 b tag $\bbMET$ region, and $\ttbar$ in all others). The systematic uncertainties are categorized as affecting either the normalization or the shape of the $\ptmiss$ distribution.  For shape uncertainties, the ranges quoted give the uncertainty in the yield for the lowest $\ptmiss$ bin and for the highest $\ptmiss$ bin.  Sources of systematic uncertainties that are common across channels are considered to be fully correlated in the channel combination fit.}
\label{tab:syst}
  \centering
     \resizebox{\textwidth}{!}{
      \begin{tabular}{l||c|c|c|c|c|c|c|c}
        \multicolumn{9}{c}{} \\
        \multicolumn{9}{c}{ Normalization uncertainties (\%)} \\
        \multicolumn{9}{c}{} \\
        \hline
        \multirow{2}{*}{Uncertainty}                           & Dileptonic             & Dileptonic               & Dileptonic                 & \semileptonic              & All-hadronic                          & All-hadronic                       & 1 b tag                            & 2 b tag            \\
                                        & $\tteeMET$ & $\ttemMET$ & $\ttmmMET$ & $\tteormMET$& $\ttoneRTTMET$   & $\tttwoRTTMET$  & $\bbMET$                      & $\bbMET$      \\
        \hline
        Integrated luminosity                            & \multicolumn{3}{c|}{2.7}                                                       & 2.7                       & \multicolumn{2}{c|}{2.7}                                           & \multicolumn{2}{c}{2.7}                                          \\
        \hline
        Pileup                               & \multicolumn{3}{c|}{0.2}                                                       & 1.4                       & \multicolumn{2}{c|}{0.4}                                           & \multicolumn{2}{c}{0.6}                                          \\
        \hline
   $\WZjets$ heavy flavor fraction            & \multicolumn{3}{c|}{\NA}                                                         &  20                       & \multicolumn{2}{c|}{20}                                            & \multicolumn{2}{c}{\NA}                                            \\
        \hline
     Drell--Yan bkg. normalization                  &  64                   & \NA                         & 43                         &     \NA                     & \multicolumn{2}{c|}{\NA}                                             & \multicolumn{2}{c}{\NA}                                            \\
        \hline
    Single t bkg. normalization                  & \multicolumn{3}{c|}{20}                                                        &    20                     & \multicolumn{2}{c|}{20}                                            & \multicolumn{2}{c}{15}                                           \\
        \hline
    Multijet bkg. normalization                         & \multicolumn{3}{c|}{\NA}                                                         &     \NA                     & \multicolumn{2}{c|}{100}                                           & \multicolumn{2}{c}{50}                                           \\
        \hline
    Misid. lepton normalization               & 200                   & 30                        & 48                         &     \NA                     & \multicolumn{2}{c|}{\NA}                                             & \multicolumn{2}{c}{\NA}                                            \\
        \hline
    RTT efficiency                    & \multicolumn{3}{c|}{\NA}                                                         &     \NA                     & \multicolumn{2}{c|}{4}                                             & \multicolumn{2}{c}{\NA}                                            \\
        \hline
       b tagging efficiency                       & \multicolumn{3}{c|}{2.2}                                                       &   2.9                     &  7.5                              & 2.3                            & \multicolumn{2}{c}{12}                                           \\
        \hline
       Lepton efficiency                      & \multicolumn{3}{c|}{4}                                                         &  2                        & \multicolumn{2}{c|}{\NA}                                             & \multicolumn{2}{c}{\NA}                                            \\
        \hline
   \ptmiss trigger efficiency                 & \multicolumn{3}{c|}{\NA}                                                         &     \NA                     & \multicolumn{2}{c|}{2}                                             & \multicolumn{2}{c}{0.3}                                          \\
        \hline
   Lepton trigger efficiency                  & \multicolumn{3}{c|}{1}                                                         &  2                        & \multicolumn{2}{c|}{\NA}                                             & \multicolumn{2}{c}{\NA}                                            \\
        \hline
              \multicolumn{9}{c}{} \\
        \multicolumn{9}{c}{ Shape uncertainties (\%)} \\
                \multicolumn{9}{c}{} \\
        \hline
        \multirow{2}{*}{Uncertainty}                    & Dileptonic             & Dileptonic               & Dileptonic                 & \semileptonic              & All-hadronic                          & All-hadronic                       & 1 b tag                           & 2 b tag           \\
                                 & $\tteeMET$ & $\ttemMET$ & $\ttmmMET$ & $\tteormMET$ & $\ttoneRTTMET$   & $\tttwoRTTMET$  & $\bbMET$                      & $\bbMET$     \\
        \hline
   PDFs                                 & \multicolumn{3}{c|}{1.6 -- 2.2}                                     &  1.8 -- 2.9    & 1.6 -- 4.9             & 1.9 -- 3.4          & 1.0 -- 2.0                   & 0.2 -- 0.8 \\
        \hline
   Jet energy scale                    & \multicolumn{3}{c|}{0.6 -- 14}                                      &  13 -- 21      & 10 -- 75               & 11 -- 24            & \multicolumn{2}{c}{1.3 -- 2.6}                       \\
        \hline
   Top quark \pt reweighting                    & \multicolumn{3}{c|}{0.9 -- 17}                                      &  10 -- 12      & 13 -- 23               & 15 -- 18            & \multicolumn{2}{c}{\NA}                                           \\
        \hline
   Diboson $\muR,\:\muF$             & \multicolumn{3}{c|}{4.1 -- 12}                                      &  12 -- 15     & 10 -- 18                & 3.2 -- 23           & \multicolumn{2}{c}{15 -- 15}                         \\
        \hline
   $\ttbar+{\Z/\PW}\gamma$ $\muR,\:\muF$        & \multicolumn{3}{c|}{11  -- 25}                                      &  14 -- 26      & 11 -- 25               & 10 -- 15            & \multicolumn{2}{c}{\NA}                                           \\
        \hline
   $\ttbar$ $\muR,\:\muF$          & \multicolumn{3}{c|}{13  -- 23}                                      &  19 -- 38      & 13 -- 25               & 22 -- 37            & \multicolumn{2}{c}{\NA}                                           \\
        \hline
   $\WZjets$ $\muR$                    & \multicolumn{3}{c|}{\NA}                                                         &  7.8 -- 8.8    & \multicolumn{2}{c|}{6.9 -- 10}                         & \multicolumn{2}{c}{4.4 -- 5.6}                       \\
        \hline
   $\WZjets$ $\muF$                    & \multicolumn{3}{c|}{\NA}                                                         &  1.4 -- 2.6    & \multicolumn{2}{c|}{0.2 -- 3.5}                         & \multicolumn{2}{c}{2.8 -- 11}                        \\
        \hline
   $\WZjets$ EWK correction     & \multicolumn{3}{c|}{\NA}                                                         &   14 -- 20     & \multicolumn{2}{c|}{4.2 -- 14}                          & \multicolumn{2}{c}{4.8 -- 21}                        \\
        \hline
        \end{tabular}
}
\end{table*}

The following sources of uncertainty correspond to constrained normalization nuisance parameters in the fit:

\begin{itemize}
\item \textbf{Integrated luminosity:} An uncertainty of 2.7\% is used for the integrated luminosity of the data sample~\cite{CMS-PAS-LUM-15-001}.

\item \textbf{Pileup modeling:} Systematic uncertainties due to pileup modeling are taken into account by varying the total inelastic cross section used to calculate the data pileup distributions by $\pm$5\%.  Normalization differences in the range of 0.2--1.4\% result from reweighting the simulation accordingly.

\item \textbf{W/Z+heavy-flavor fraction:}  The uncertainty in the fraction of W/Z + heavy-flavor jets is assigned to account for the usage of CRs dominated by light-flavor jets in constraining the prediction of $\Wjets$ and $\Zjets$ in SRs that require b tags.  The flavor fractions for the $\Wjets$ and $\Zjets$ processes are allowed to vary independently within 20\%~\cite{CMS-PAS-SMP-12-023,Chatrchyan2014204,CMS-PAS-SMP-12-017,Chatrchyan:2014dha}.

\item \textbf{Drell--Yan background}: The uncertainties in the data-driven Drell--Yan background estimates for the dileptonic channels are 64\% ($\ee$) and 43\% ($\mumu$).  These uncertainties are dominated by the statistical uncertainties in quantities used to extrapolate yields from a region near the Z boson mass to regions away from it.  Again, these relatively large uncertainties have little effect on the sensitivity of the search.

\item \textbf{Multijet background normalization:} Uncertainties of 50--100\% (depending on the SR) are applied in the normalization of multijet backgrounds to cover tail effects that are not well modeled by the simulation.

\item \textbf{Misidentified-lepton background}: The sources of uncertainty in the misidentified-lepton background for the dileptonic search stem from the uncertainty in the measured misidentification rate, and from the statistical uncertainty of the single-lepton control sample to which the rate is applied. The uncertainties per channel are 200\% ($\ee$), 48\% ($\emu$), 30\% ($\mumu$), and are dominated by the statistical uncertainty associated with the single-lepton control sample.  Because the misidentified lepton background is small, these relatively large uncertainties do not significantly degrade the sensitivity of the search.

\item \textbf{RTT efficiency:} Jet energy scale and resolution uncertainties are propagated to the RTT efficiency scale factors by using modified shape templates in the efficiency extraction fit.  A systematic uncertainty due to the choice of parton showering scheme is estimated by comparing the efficiencies obtained with default and alternative $\ptmiss$ templates.  The default simulation is showered using {\PYTHIA 8.205}, which implements dipole-based parton showering.  The alternative templates are derived from simulated events that are showered with {\HERWIG}~\cite{Bahr:2008pv}, which uses an angular-ordered shower model.  Overall, statistical plus systematic uncertainties of 6, 3, and 3\% are assigned for the hadronic tag, hadronic mistag, and nonhadronic mistag scale factors, respectively.  These correspond to an overall normalization uncertainty for the $\ttMET$ SRs of 4\%.

\item \textbf{b tagging efficiency:} The b tagging efficiency and its uncertainty are measured using independent control samples.  Uncertainties from gluon splitting, the b quark fragmentation function, and the selections used to define the control samples are propagated to the efficiency scale factors~\cite{btag}.  The corresponding normalization uncertainty ranges from 2.2 to 12\%.

\item \textbf{Lepton identification and trigger efficiency:} The uncertainty in lepton identification and triggering efficiency is measured with samples of Z bosons decaying to dielectrons and dimuons~\cite{CMS:2011aa}.  The corresponding normalization uncertainty ranges from 2 to 4\%.

\item \textbf{$\ptmiss$ trigger:} Uncertainties of 0.3--2\% (depending on the SR) are associated with the efficiency scale factors of the $\ptmiss$ trigger. The efficiency of this trigger is measured using data collected with the single-lepton triggers. For values of $\ptmiss>200\GeV$, these data primarily consist of \Wjets~events.
\end{itemize}

The following sources of uncertainty correspond to constrained \ptmiss~shape nuisance parameters in the fit:
\begin{itemize}

\item \textbf{PDF uncertainties:}  Uncertainties due to the choice of PDFs are estimated by reweighting the samples with the ensemble of PDF replicas provided by NNPDF3.0 \cite{Butterworth:2015oua}.  The standard deviation of the reweighted $\ptmiss$ shapes is used as an estimate of the uncertainty.

\item \textbf{Jet energy scale:} Reconstructed jet four-momenta in the simulation are simultaneously varied according to the uncertainty in the jet energy scale~\cite{2011JInst...611002C}. Jet energy scale uncertainties are coherently propagated to all observables including $\ptmiss$.

\item \textbf{Top quark $\pt$ reweighting:} Differential measurements of top quark pair production show that the measured $\pt$ spectrum of top quarks is softer than that of simulation. Scale factors to cover this effect have been derived in previous CMS measurements~\cite{Chatrchyan:2012saa} and are applied to all simulated SM $\ttbar$ samples by default.  The uncertainty in the top quark $\pt$ spectrum is estimated from a comparison with the spectrum obtained without reweighting.

\item \textbf{Higher-order QCD corrections:} The uncertainties due to missing higher-order QCD corrections in the LO samples are estimated by generating alternative event samples in which the factorization and renormalization scale parameters ($\muF{},\muR$) are simultaneously increased or decreased by a factor of two.  These uncertainties are correlated across the bins of the $\ptmiss$ distribution.  Uncertainties in the NLO K-factors applied to $\Wjets$ and $\Zjets$ simulation are determined by recalculating the K-factor with $\muF$ and $\muR$ independently varied by a factor of two up or down.

\item \textbf{EWK corrections:}  Uncertainties in the K-factors applied to $\Wjets$ and $\Zjets$ simulation from missing higher-order EWK corrections are estimated by taking the difference in results obtained with and without the EWK correction applied.

\item \textbf{Simulation statistics:}  Shape uncertainties due to the limited sizes of the simulated signal and background samples are included via the method of Barlow and Beeston~\cite{BARLOW1993219,Conway-PhyStat}.  This approach allows each bin of the $\ptmiss$ distributions to independently fluctuate according to Poisson statistics.

\end{itemize}

\section{Results and interpretation}\label{sec:results}
Separate signal strength parameters are first determined from fits to each of the $\bbMET$ and $\ttMET$ channels.  These fits use the predicted cross sections and $\ptmiss$ shapes from the LHC DMF signal models with $\gq = g_{\chi} = 1$.  The fits result in independent upper limits on signal yields for the $\bbDM$ and $\ttDM$ processes, which are reported in Section~\ref{sec:results_separate}.

Next, all SRs and CRs are simultaneously fit under the hypothesis of combined $\ttDM$ and $\bbDM$ contributions.  In this case, a single signal strength parameter is used, which results in a combined best fit estimate of the $\ttDM$ and $\bbDM$ signal yields.  Again, cross section predictions for  $\ttDM$ and $\bbDM$ assume $\gq=g_{\chi}=1$.  Results from this fit are reported in Section~\ref{sec:results_combined}.

The most interesting DM scenarios to explore at the LHC involve on-shell mediator decays to $\XX$, which corresponds to $m_{\phi/\mathrm{a}} > 2m_{\chi}$.  Kinematic variables and cross sections are independent of $m_{\chi}$ in this regime~\cite{Abercrombie:2015wmb}.  The $m_{\chi} < 10\GeV$ region is of particular interest because of the strong phenomenological and theoretical motivations for low-mass DM~\cite{Lin:2011gj} and the relative strength of collider experiments in this mass range~\cite{Bauer:2013ihz}.  For these reasons, the DM mass has been fixed to $m_{\chi}=1\GeV$ in all signal extraction fits. The results obtained with $m_{\chi}=1\GeV$ are valid for other values of $m_{\chi} < m_{\phi/\mathrm{a}}/2$ provided they are not too near the kinematic threshold.

\subsection{Individual search results}\label{sec:results_separate}
\begin{figure*}[h!tbp]
\centering
  \includegraphics[width=0.49\textwidth]{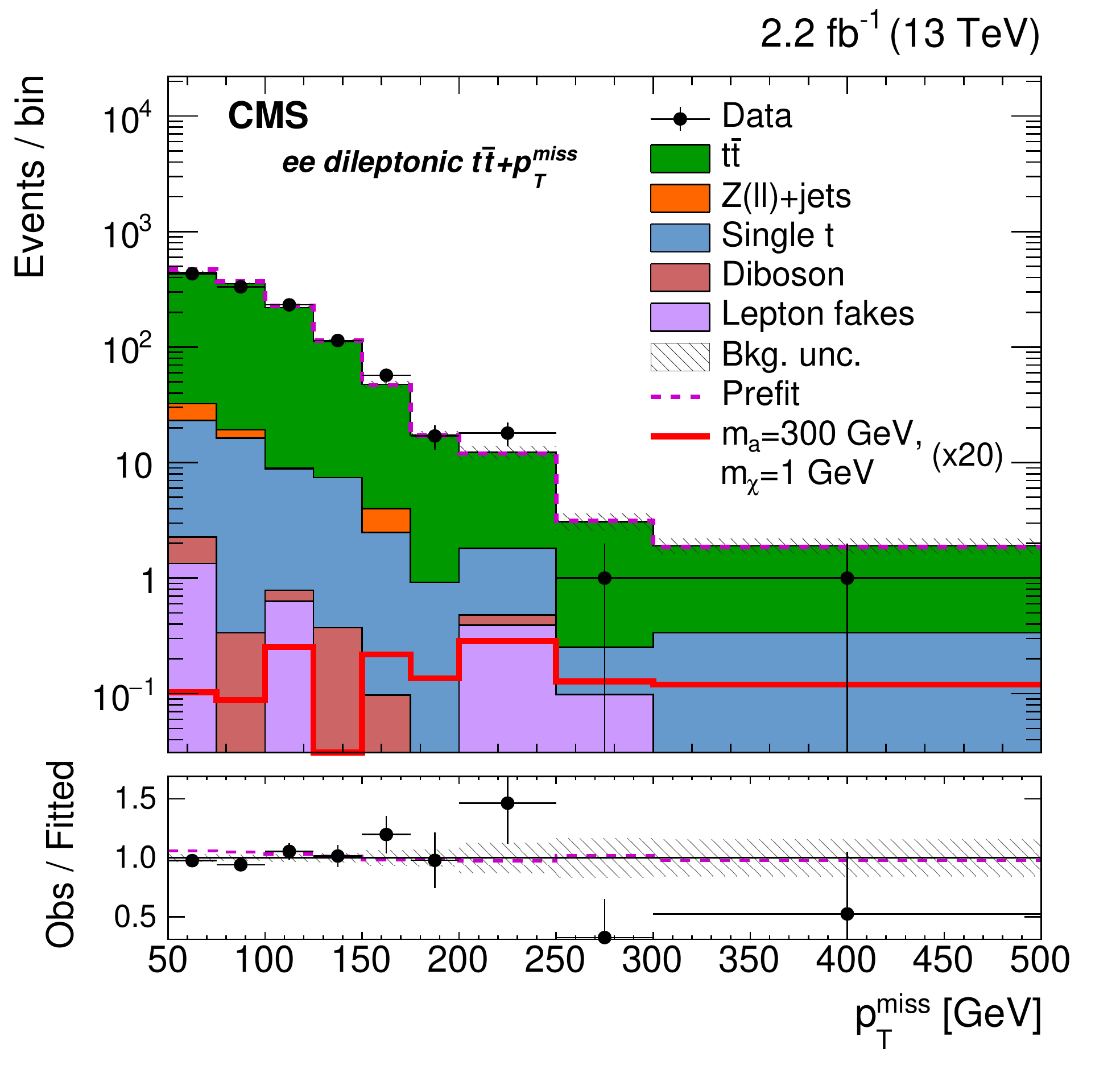}
  \includegraphics[width=0.49\textwidth]{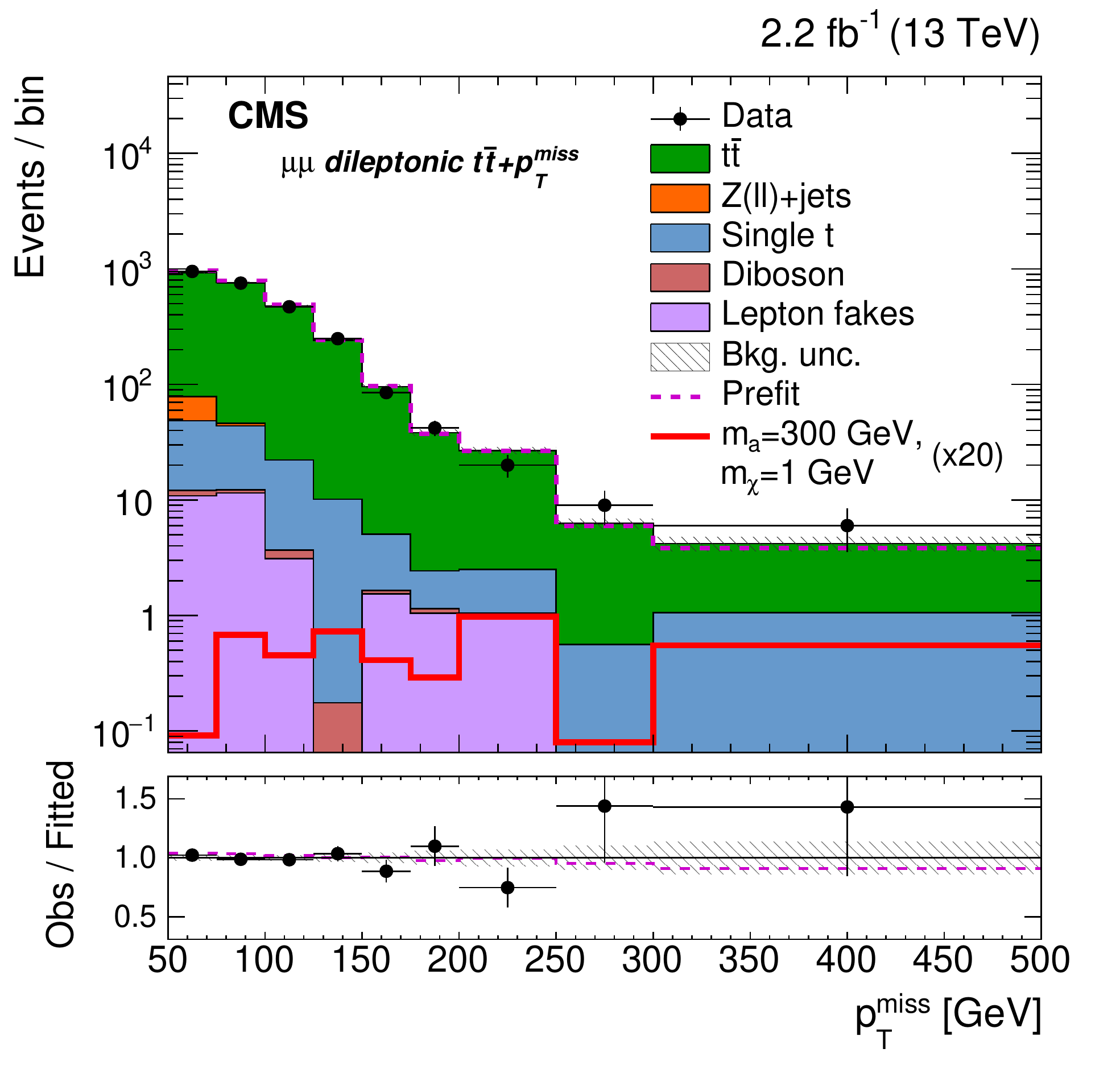}
  \includegraphics[width=0.49\textwidth]{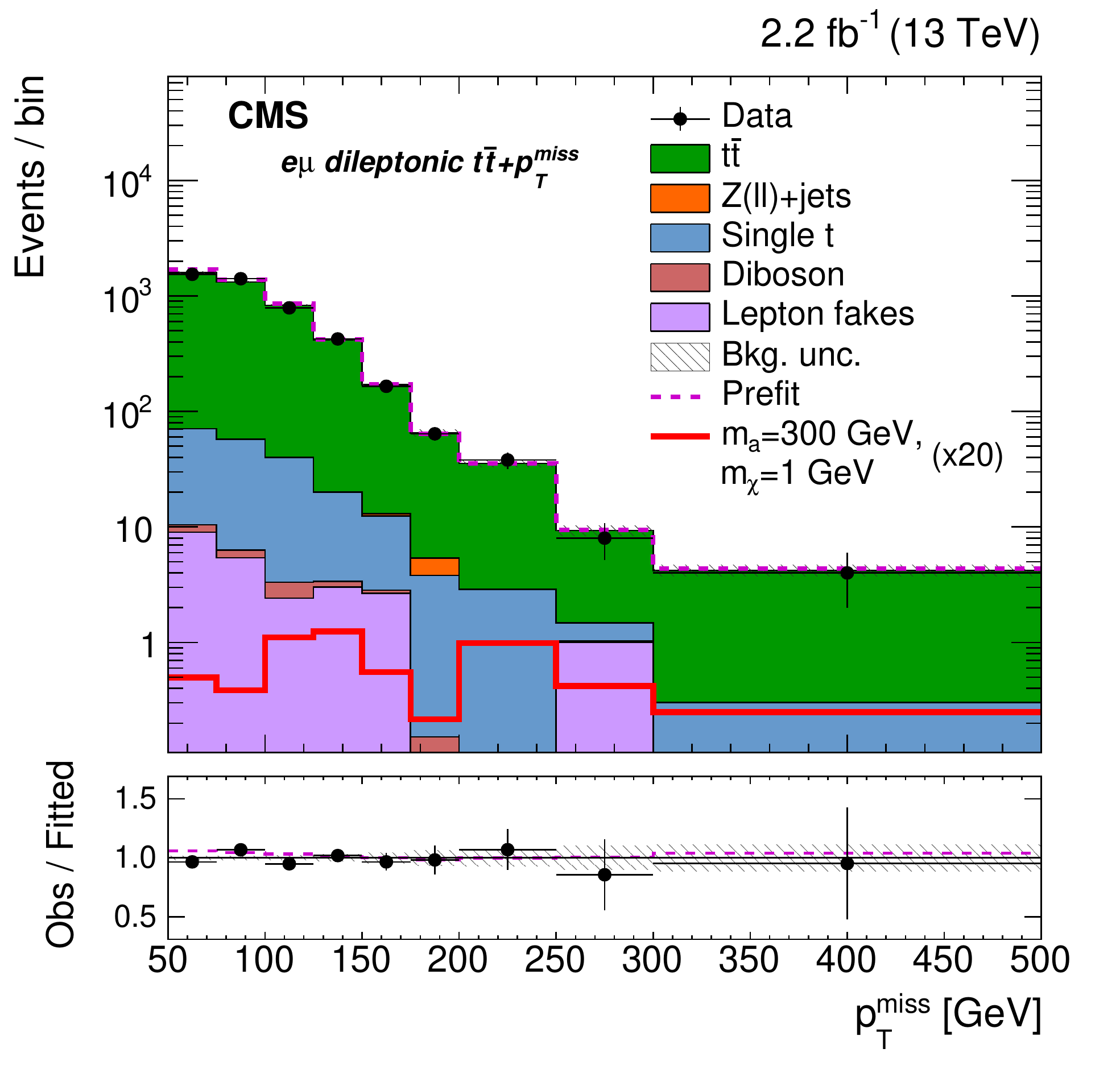}
  \includegraphics[width=0.49\textwidth]{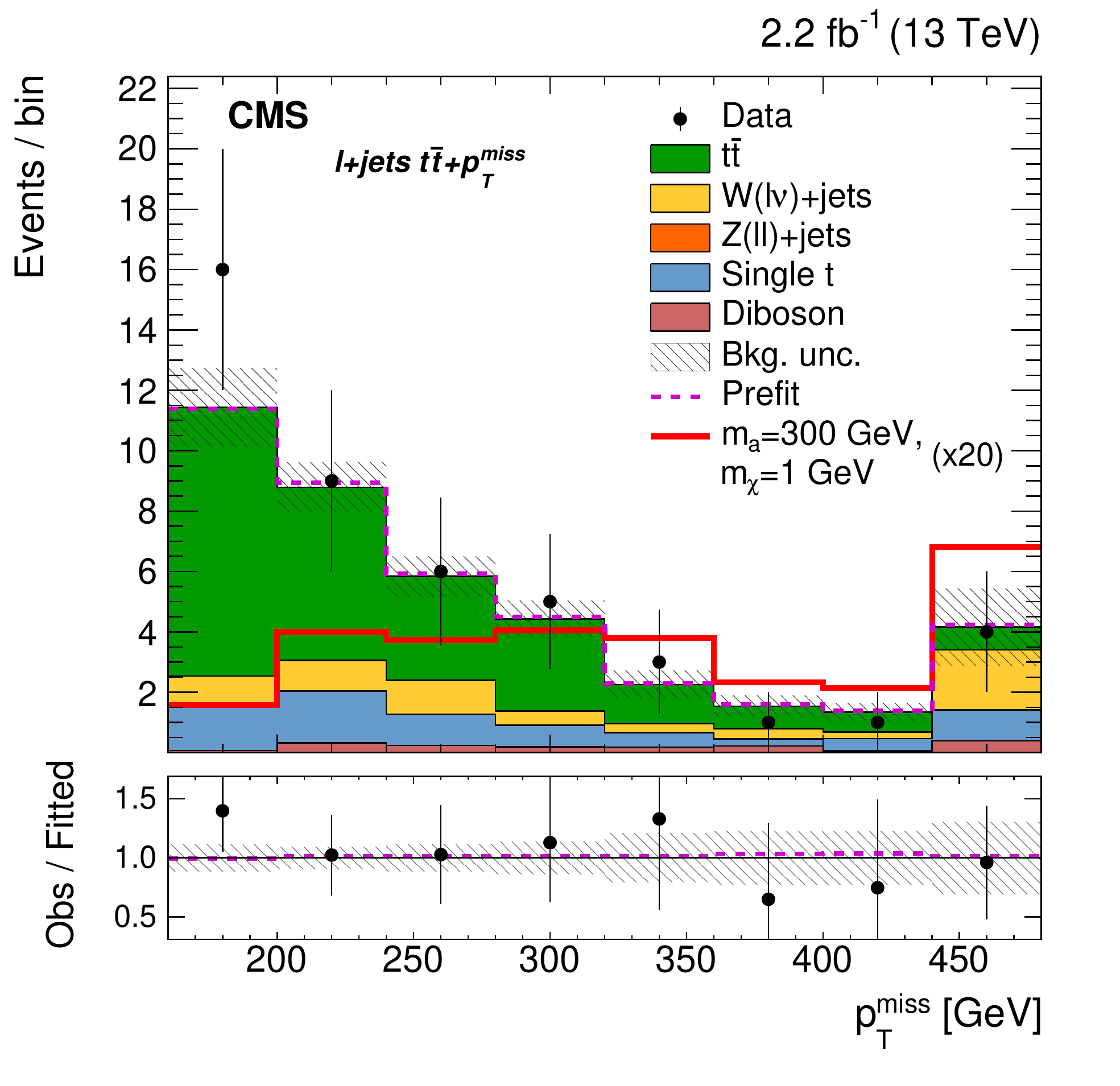}
  \caption{The $\ptmiss$ distributions in the following signal regions: dileptonic $\ttMET$ in the $\ee$ signal region (upper left), in the $\mumu$ region (upper right), in the $\emu$ region (lower left), and in \semileptonic $\ttMET$ region (lower right).  The $\ptmiss$ distributions of background correspond to background-only fits to the individual $\ttMET$ signal regions and associated background control regions.  The prefit $\ptmiss$ distribution of an example signal (pseudoscalar mediator, $m_{\mathrm{a}} = 300\GeV$ and $m_{\chi} = 1\GeV$) is scaled up by a factor of 20.  The last bin contains overflow events.  The lower panels of each plot show the ratio of observed data to fitted background.  The uncertainty bands shown in these panels are the fitted values, and the magenta lines correspond to the ratio of prefit to fitted background expectations.}
  \label{fig:postfit_ptmiss_separated_PS_lep}
\end{figure*}

\begin{figure*}[h!tbp]
\centering
  \includegraphics[width=0.49\textwidth]{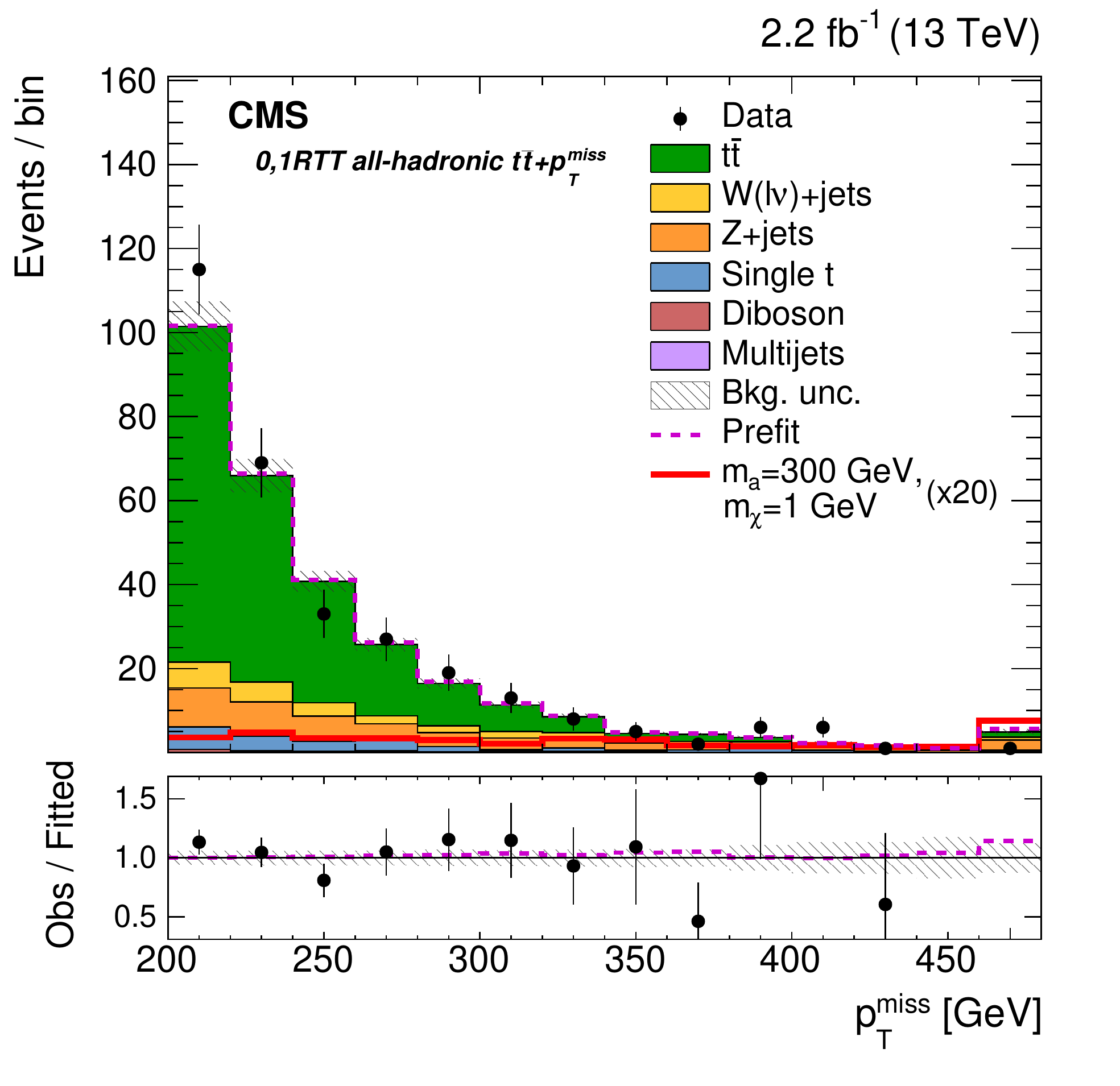}
  \includegraphics[width=0.49\textwidth]{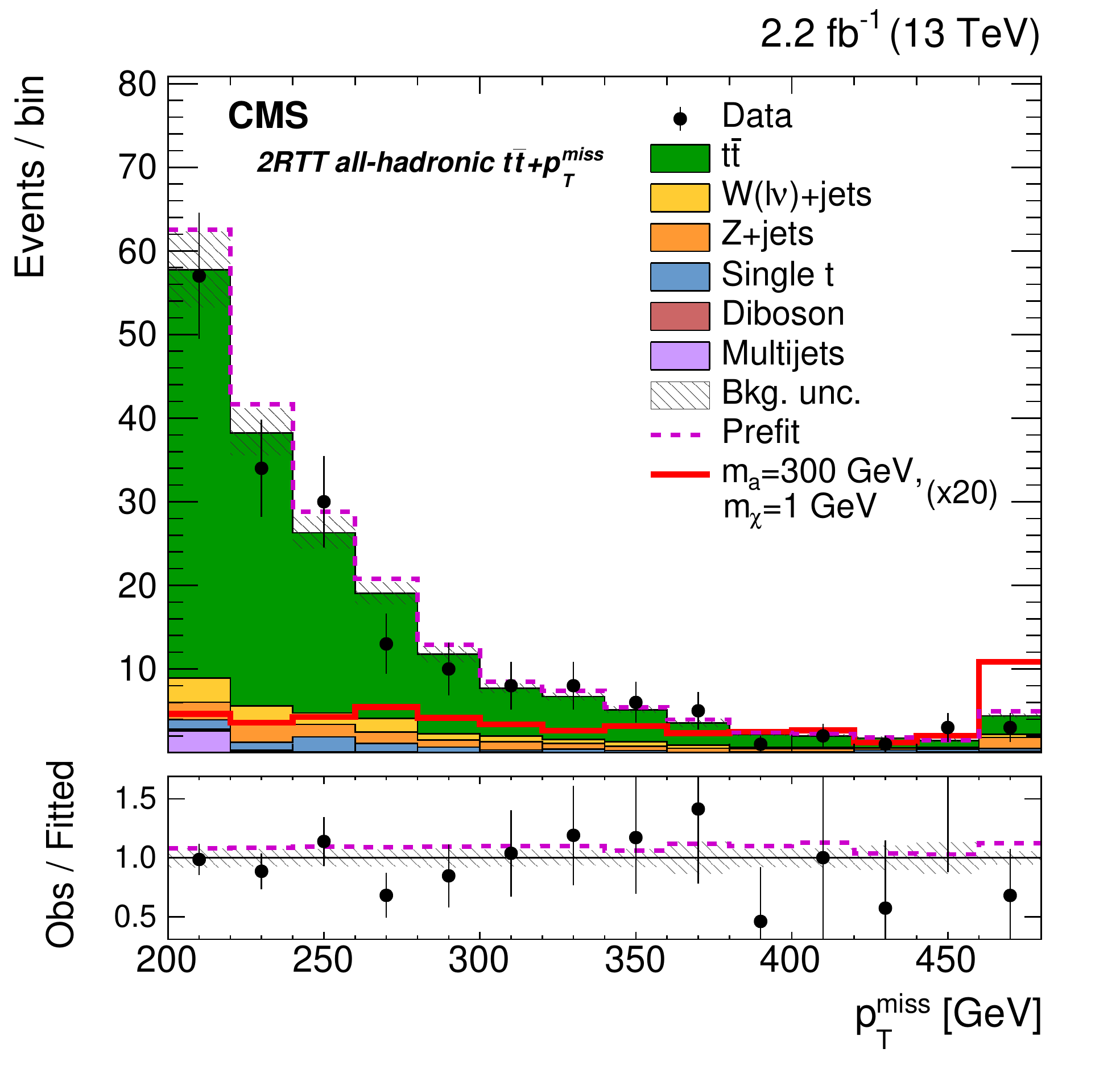}
  \includegraphics[width=0.49\textwidth]{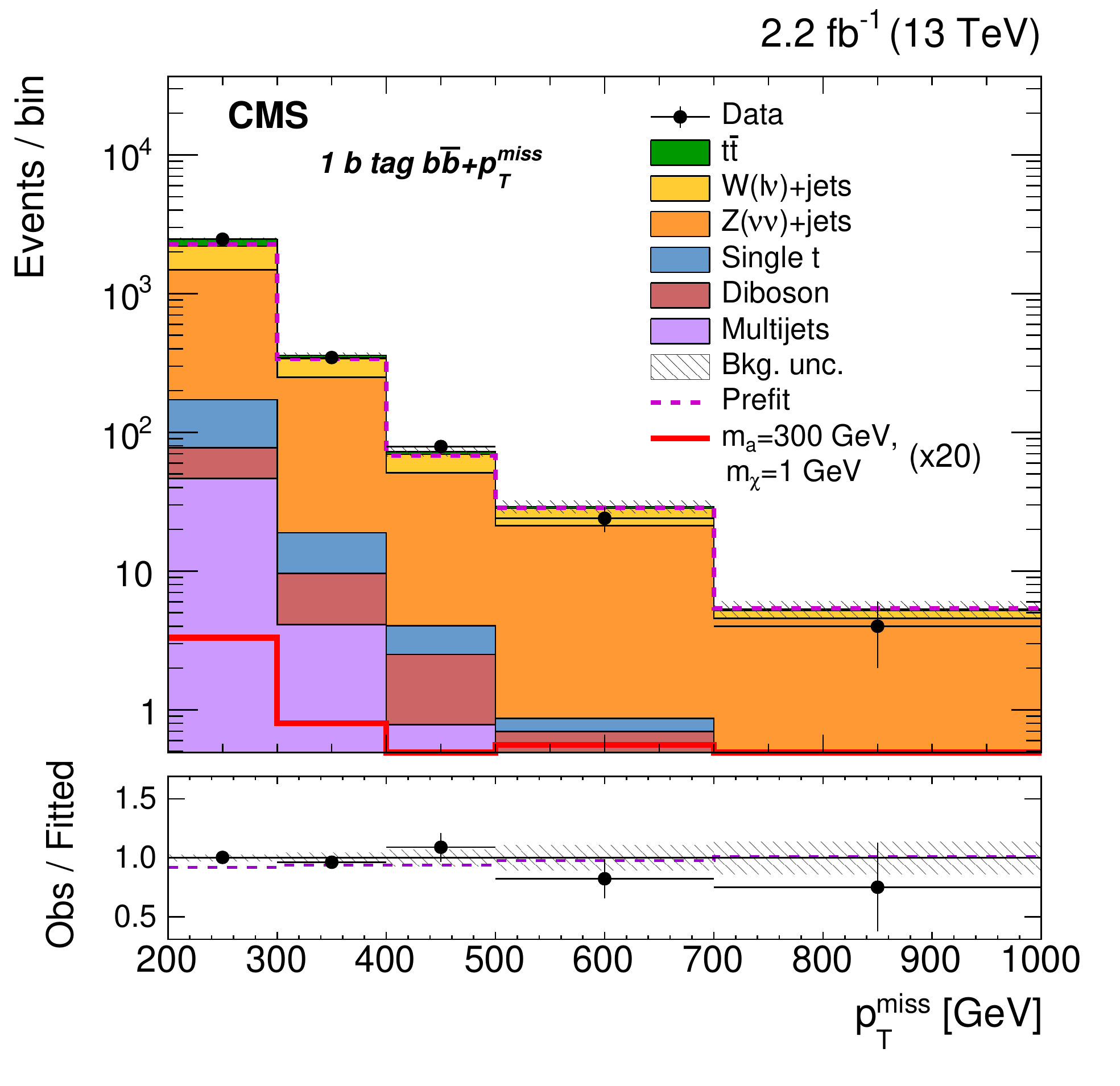}
  \includegraphics[width=0.49\textwidth]{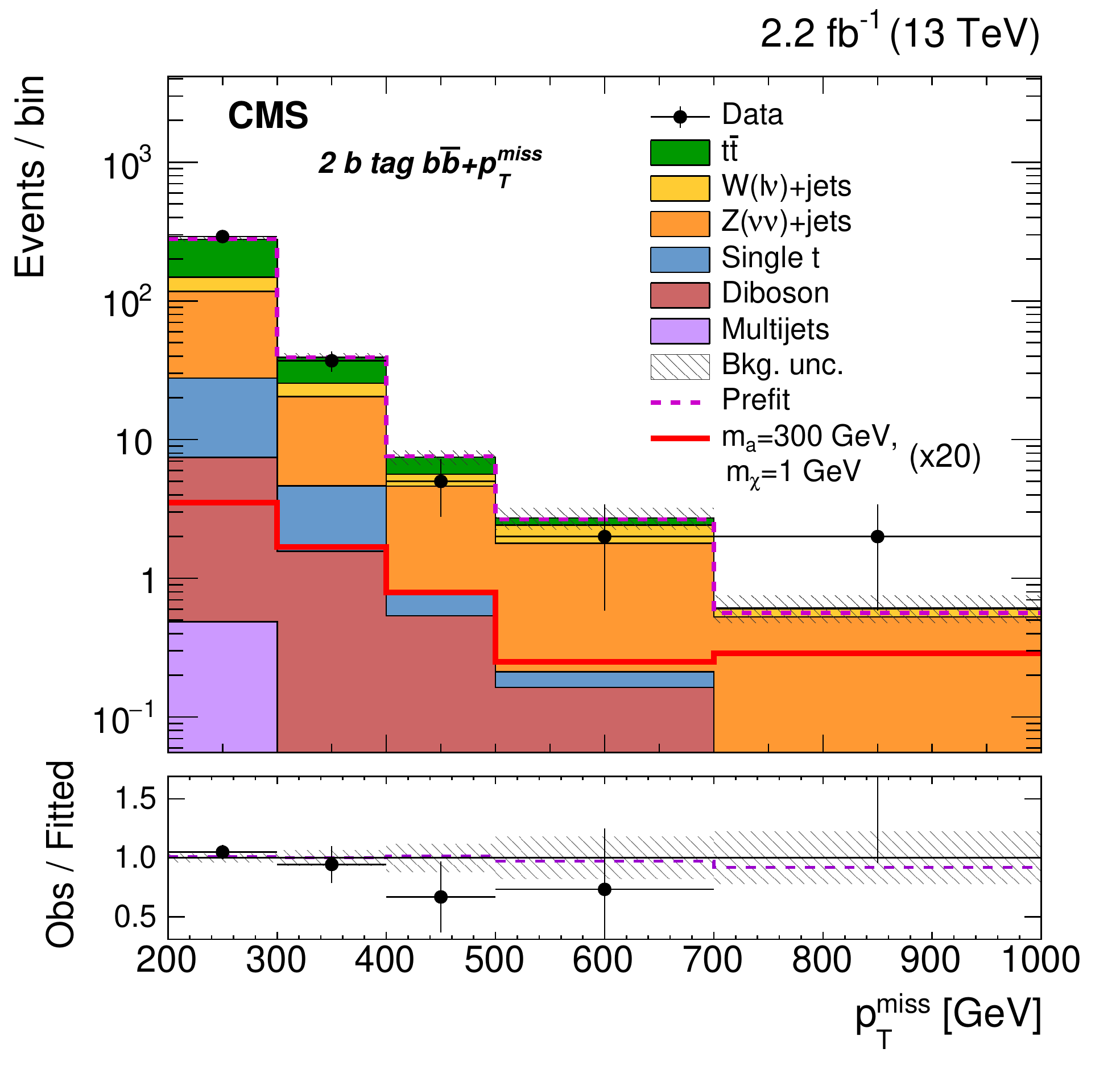}
  \caption{The $\ptmiss$ distributions in the following signal regions: all-hadronic $\ttMET$ with 0 or 1 RTTs (upper left), all-hadronic $\ttMET$ with 2 RTTs (upper right), $\bbMET$ with 1 b tag (lower left), and $\bbMET$ with 2 b tags (lower right).  The $\ptmiss$ distributions of background correspond to background-only fits to the individual $\ttMET$ and $\bbMET$ signal regions and associated background control regions.  The prefit $\ptmiss$ distribution of an example signal (pseudoscalar mediator, $m_{\mathrm{a}} = 300\GeV$ and $m_{\chi} = 1\GeV$) is scaled up by a factor of 20.  The last bin contains overflow events.  The lower panels of each plot show the ratio of observed data to fitted background.  The uncertainty bands shown in these panels are the fitted values, and the magenta lines correspond to the ratio of prefit to fitted background expectations.}
  \label{fig:postfit_ptmiss_separated_PS_nolep}
\end{figure*}

Table~\ref{tab:postfit_yields_separated_PS} provides the background yields in the SRs obtained from background-only fits to the $\bbMET$ and individual $\ttMET$ search channels.  Relative nuisance parameter shifts --- defined as $(\text{p}_{\text{fit}} - \text{p}_{\text{prefit}}) / \sigma_{\text{p}}$, where $\text{p}$ represents the parameter value and $\sigma_{\text{p}}$ its fit uncertainty --- do not indicate any particular tension in these fits.  The largest shifts correspond to the nuisance parameters for the EWK correction for the $\Wjets$ and $\Zjets$ processes in the $\bbMET$ channel (+0.8), to the $\mu_F\:,\mu_R$ scale uncertainty in the $\ttbar$ process in the \semileptonic $\ttMET$ channel (+0.6), and to the lepton efficiency in the all-hadronic $\ttMET$ channel ($-1.9$). The nuisance parameter shifts account for residual mismodeling of the yields by the simulation in the background-enriched regions.  The background-only fitted $\ptmiss$ distributions in the eight SRs are shown in Figs.~\ref{fig:postfit_ptmiss_separated_PS_lep} and~\ref{fig:postfit_ptmiss_separated_PS_nolep}.

\begin{table*}[h!]
  \topcaption{Fitted background yields for a background-only hypothesis in the $\ttMET$ and $\bbMET$ signal regions.  The yields are obtained from separate fits to the $\bbMET$ and individual $\ttMET$ search channels.  Prefit yields for DM produced via a pseudoscalar mediator with mass $m_{\mathrm{a}}=50\GeV$ and a scalar mediator with mass $m_{\phi}=100\GeV$ are also shown.  Mediator couplings are set to $\gq=g_{\chi}=1$, and a DM particle of mass $m_{\chi}=1\GeV$ is assumed.  Uncertainties include both statistical and systematic components. }
  \label{tab:postfit_yields_separated_PS}
   \centering
    \def\arraystretch{1.3}
    \resizebox{\textwidth}{!}{
      \begin{tabular}{c||c|c|c|c|c|c|c|c}
\hline
        \multirow{2}{*}{Channel}     & \multicolumn{3}{c|}{Dileptonic} & \semileptonic  & \multicolumn{2}{c|}{All-hadronic} & \multicolumn{2}{c}{\multirow{2}{*}{$\bbMET$}} \\
                                     & \multicolumn{3}{c|}{$\ttMET$} &  $\ttMET$  & \multicolumn{2}{c|}{$\ttMET$}  & \multicolumn{2}{c}{} \\
        \hline
        Signal Region                               & $\ee$ & $\emu$ & $\mumu$ & $\eorm$ & 0,1 RTT   & 2 RTT  & 1 b tag  & 2 b tags \\
        \hline
        \hline
        $\ttbar$                              & $1133 \pm 29$          & $4228 \pm 73$            & $2412 \pm 51$              & $24.6 \pm 2.2$            &   $203 \pm 18$                    &  $152 \pm 13$                  &  $284 \pm 28$                           &   $145 \pm 11$                          \\
        $\Wjets$                              &        \NA               &        \NA                 &        \NA                   &  $6.4 \pm 1.6$            &  $23.1 \pm 4.5$                   & $11.9 \pm 1.3$                 &  $829 \pm 59$                           &  $38.5 \pm 5.5$                         \\
        $\Zjets$                              &   $14 \pm 12$          &  $2.5 \pm 4.7$           &   $32 \pm 15$              & $0.10 \pm 0.04$           &    $44 \pm 11$                    & $13.0 \pm 1.3$                 & $1613 \pm 64$                           & $110.7 \pm 6.7$                         \\
        Single t                              &   $57 \pm 12$          &  $182 \pm 36$            &  $104 \pm 22$              &  $7.0 \pm 2.0$            &  $19.1 \pm 2.0$                   &  $7.3 \pm 1.4$                 &  $105 \pm 16$                           &  $23.6 \pm 4.0$                         \\
        Diboson                              &   $2.0 \pm 0.4$        &  $4.0 \pm 0.6$           &  $3.1 \pm 0.5$             &  $1.7 \pm 0.4$            &  $3.3 \pm 0.3$                    & $1.0 \pm 0.3$                  & $38.7 \pm 6.6$                          &   $9.2 \pm 1.6$                         \\
        Multijets                             &       \NA                &        \NA                 &        \NA                   &       \NA                   & $0.10 \pm 0.08$                   &  $2.9 \pm 2.2$                 &   $52 \pm 22$                           &  $0.5 \pm 0.2$                        \\
    Misid. lepton                             &  $2.5 \pm 7.7$         & $24 \pm 11$              & $29.0 \pm 8.7$             &       \NA                   &        \NA                          &       \NA                        &        \NA                                &         \NA                               \\
        \hline
        Background                            & $1208 \pm 32$          & $4439 \pm 71$            & $2580 \pm 52$              & $39.8 \pm 3.4$            &   $293 \pm 21$                    &  $188 \pm 12$                  & $2922 \pm 77$                           &   $327 \pm 12$                          \\
        \hline
        Data                                  & 1203                   & 4436                     & 2585                       & 45                        & 305                               & 181                            & 2919                                    & 337                                     \\
        \hline
        \multicolumn{9}{c}{  }\\
        \multicolumn{9}{c}{ $m_{\mathrm{a}}=50\GeV$ }\\
        \hline
        $\ttDM$             & $1.19 \pm 0.37$        & $3.48 \pm 0.73$          & $1.62 \pm 0.36$            & $5.9 \pm 1.0$             &   $7.5 \pm 1.5$                   &  $8.4 \pm 1.8$                 & $1.21 \pm 0.38$                         & $1.34 \pm 0.34$                         \\
        $\bbDM$             & $0 \pm 0$              & $0 \pm 0$                & $0 \pm 0$                  &   $0 \pm0 $               & $0.01 \pm 0.05$                 &    $0 \pm0 $                   & $3.44 \pm 0.94$                         & $0.55 \pm 0.22$                         \\
        \hline
        \multicolumn{9}{c}{ $m_{\phi}=100\GeV$ }\\
        \hline
        $\ttDM$         & $1.27 \pm 0.49$        & $6.3 \pm 1.1$            & $2.51 \pm 0.76$            & $4.44 \pm 0.95$           &   $7.3 \pm 2.0$                   & $10.2 \pm 3.1$                 & $2.22 \pm 0.53$                         & $2.11 \pm 0.64$                         \\
        $\bbDM$         & $0 \pm 0$              & $0 \pm 0$                & $0 \pm 0$                  &   $0 \pm 0$               &  $0.16 \pm 0.16$                  & $0.04 \pm 0.14$                & $2.21 \pm 0.66$                         & $0.49 \pm 0.15$                         \\
        \hline
        \end{tabular}
        }
 \end{table*}


The fitted background-only $\ptmiss$ distributions of the individual search channels are assessed using the likelihood ratio for the saturated model, which provides a generalization of the $\chi^{2}$ goodness-of-fit test~\cite{Baker1984437,Lindsey}.  Pseudodata are generated from the fitted MC yields to determine the distribution of the likelihood ratio.  The {$p$}-values obtained are larger than 0.5 for each channel except for the all-hadronic $\ttMET$ channel, for which a low $p$-value of 0.01 is determined.  This value appears to result from the scatter in the 0,1RTT CRs.  No significant excess in the individual search channels is observed.

\begin{figure}[h!tb]
\centering
  \includegraphics[width=0.49\textwidth]{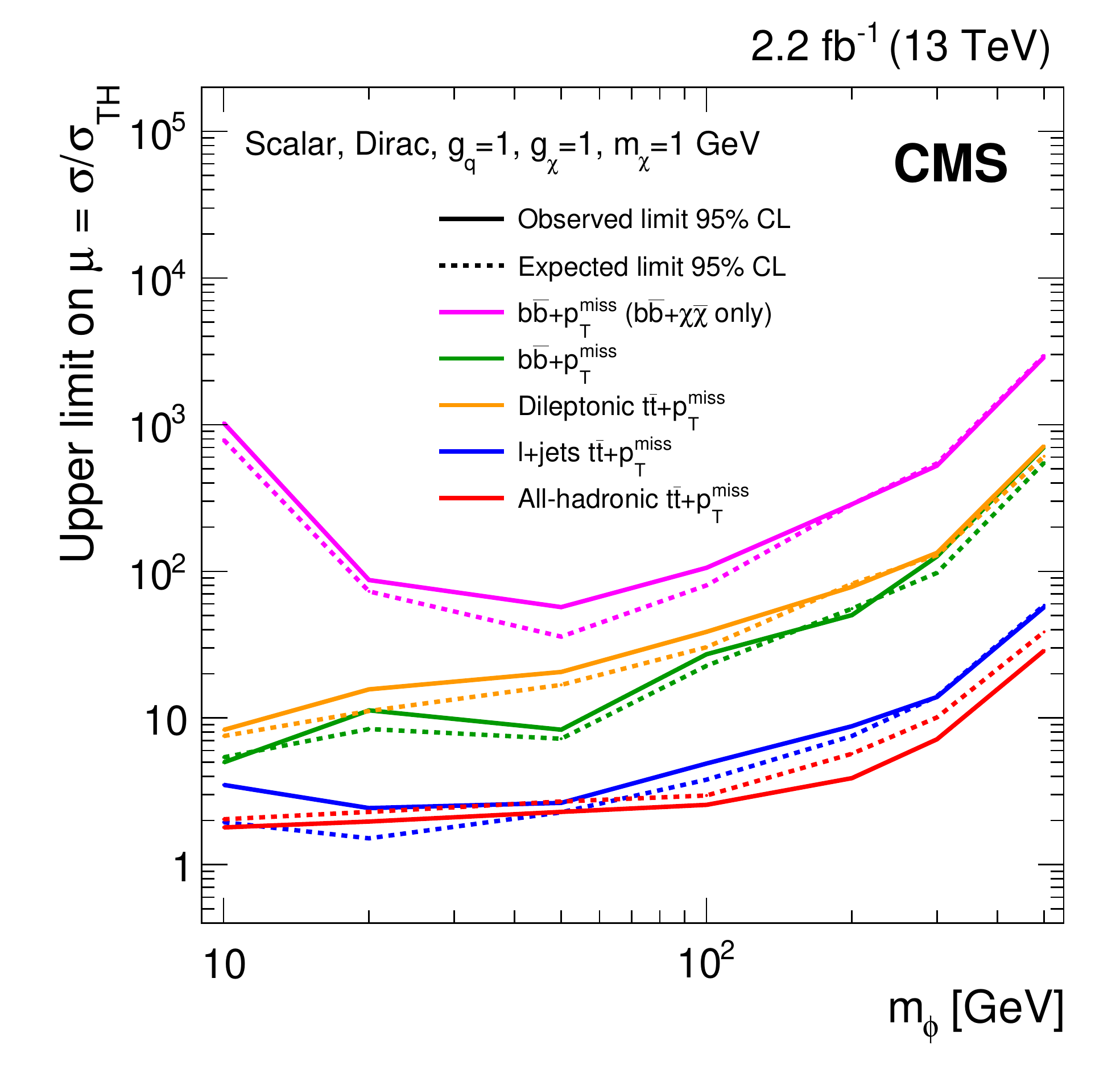}
  \includegraphics[width=0.49\textwidth]{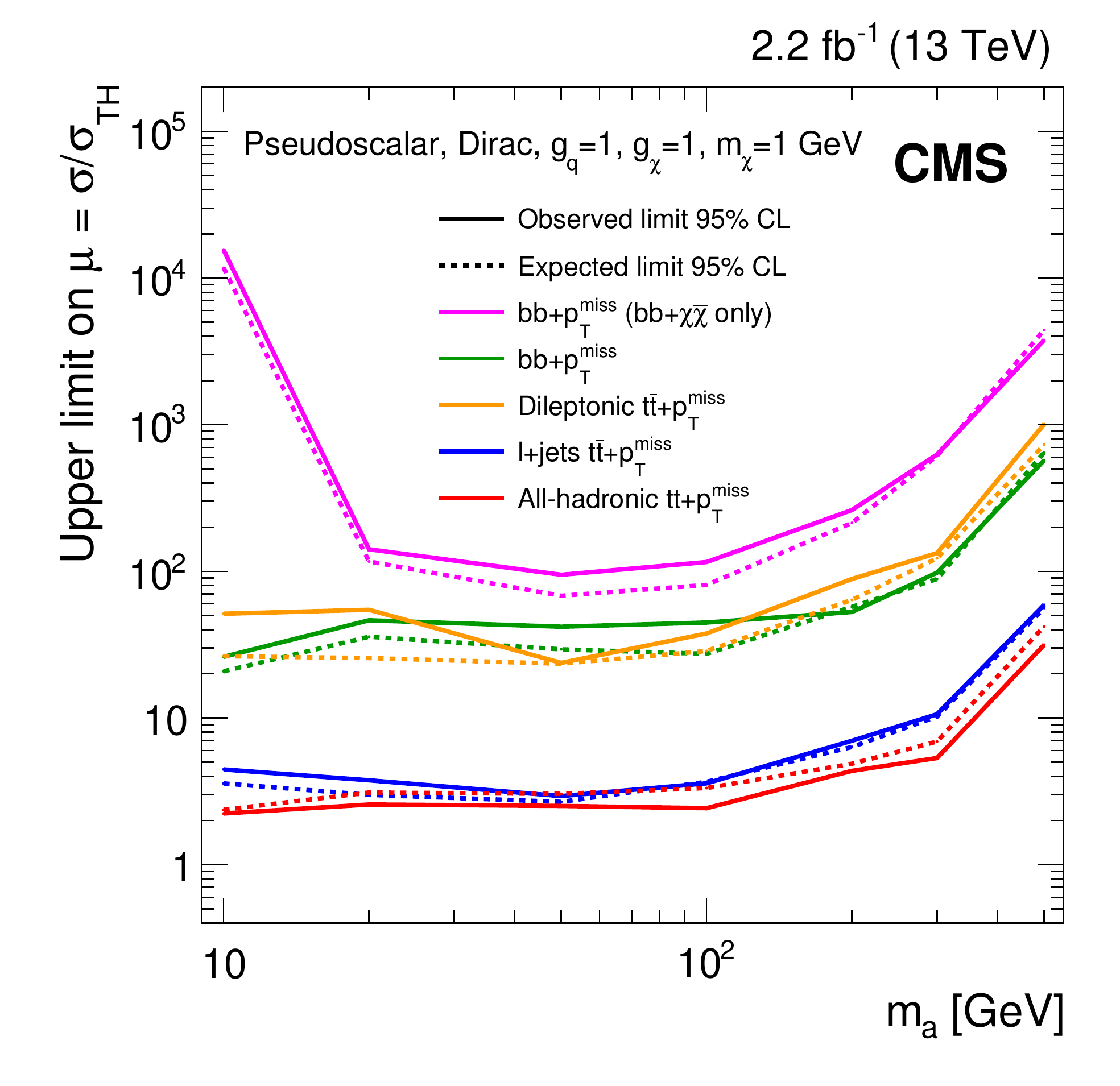}
  \caption{The ratio ($\mu$) of 95\% CL upper limits on the $\bbDM$ and $\ttDM$ cross sections to simplified model expectations.  The limits are obtained from fits to the individual $\bbMET$ and $\ttMET$ search channels for the hypothesis of a scalar mediator (\cmsLeft) or a pseudoscalar mediator (\cmsRight).  A fermionic DM particle with a mass of $1\GeV$ is assumed in both panels.  Mediator couplings correspond to $\gq=g_{\chi}=1$.}
  \label{fig:limits_separated}
\end{figure}

Upper limits are set on the $\bbDM$ and $\ttDM$ production cross sections.  The limits are calculated using a modified frequentist approach (CLs) with a test statistic based on the profile likelihood in the asymptotic approximation~\cite{JUNK1999435,cls,Cowan2011}.  For each signal hypothesis, 95\% confidence level (CL) upper limits on the signal strength parameter $\mu$ are determined.  Tables~\ref{tab:limits_separated_S} and~\ref{tab:limits_separated_PS} list the expected limits on $\mu$ obtained for various signal hypotheses.  Figure~\ref{fig:limits_separated} shows the expected and observed limits on $\mu$ as a function of the mediator mass for $m_{\chi}=1\GeV$.

The all-hadronic and \semileptonic $\ttMET$ channels provide the highest sensitivity to the $\ttDM$ process for all mediator masses considered.  Expected limits on the $\ttDM$ process from the $\bbMET$ channel are comparable with those of the dileptonic $\ttMET$ channel.  The only relevant search channel for the $\bbDM$ process is $\bbMET$, from which observed upper limits of $\mu \geq 26$ are obtained for the pseudoscalar mediator hypothesis (see Table~\ref{tab:limits_separated_PS}).  The relatively weak sensitivity of the $\bbMET$ channel in the search is due, in part, to the specific signal model considered; the performance of this channel would improve in models in which the mediator couplings to up-type quarks are suppressed.

{\tolerance=1200
In all search channels, the expected sensitivity to low-mass scalar mediators is better than that for low-mass pseudoscalars.  This reflects the higher predicted cross section for the low-mass scalar, which is approximately 40 times larger than that of the pseudoscalar for a mediator mass of 10\GeV~\cite{Backovic:2015soa}.  Scalar and pseudoscalar cross sections become comparable at mediator masses of around 200\GeV and above.  The expected scalar limits therefore rise quickly with increasing mass, while the limits for the pseudoscalar mediator change less, as can be seen from Tables~\ref{tab:limits_separated_S} and~\ref{tab:limits_separated_PS}.
\par}

\begin{table*}[h!]
\centering
  \topcaption{Observed and expected 95\% CL upper limits on the ratios ($\mu$) of the observed $\ttDM$ and $\bbDM$ cross sections to the simplified model expectations.  The limits correspond to separate fits to the $\bbMET$ and individual $\ttMET$ search channels.  DM mediators with scalar couplings of $\gq=g_{\chi}=1$ are assumed.}
  \label{tab:limits_separated_S}
   \centering
\begin{tabular}{c|c|c|c|c|c|c|c|c|c|c}\hline
& \multicolumn{8}{c|}{$\mu(\ttbar + \phi \to \ttbar \XX)$} & \multicolumn{2}{c}{$\mu(\bbbar + \phi \to \bbbar \XX)$} \\
\cline{2-11}
                    & \multicolumn{2}{c|}{Dileptonic}                                 & \multicolumn{2}{c|}{\semileptonic}                             & \multicolumn{2}{c|}{All-hadronic}                                & \multicolumn{2}{c|}{\multirow{2}{*}{$\bbMET$}} & \multicolumn{2}{c}{\multirow{2}{*}{$\bbMET$}}                 \\
$m_{\phi}$, $m_{\chi}$ & \multicolumn{2}{c|}{ $\ttMET$}  & \multicolumn{2}{c|}{$\ttMET$} & \multicolumn{2}{c|}{$\ttMET$} & \multicolumn{2}{c|}{} & \multicolumn{2}{c}{}                                                \\

\cline{2-11}
[\GeVns{}] & Obs. & Exp. & Obs. & Exp. & Obs. & Exp. & Obs. & Exp. & Obs. & Exp. \\
\hline
 10,   1 & 8.3 & 7.5 & 3.5 & 2.0 & 1.8  &  2.0  & 5.0 & 5.4 &  1.0$\times10^3$ & 789               \\
 20,   1 & 16  & 11  & 2.4 & 1.5 & 2.0  &  2.3  & 12  & 8.7 &      87  &  73                       \\
 50,   1 & 21  & 17  & 2.6 & 2.3 & 2.2  &  2.7  & 9.0 & 8.6 &      57  &  36                       \\
 100,  1 & 39  & 30  & 4.9 & 3.8 & 2.5  &  3.0  & 31  & 27  &     106  &  80                       \\
 200,  1 & 78  & 82  & 8.8 & 7.5 & 3.9  &  5.7  & 55  & 61  &     287  & 287                       \\
 300,  1 & 134 & 129 &  14 &  14 & 7.2  &   10  & 136 & 105  &     525  & 544                       \\
 500,  1 & 716 & 609 &  57 &  59 &  29  &   39  & 777 & 608 &  2.9$\times10^3$ & 3.0$\times10^3$   \\
\hline
\end{tabular}

\end{table*}

\begin{table*}[h!tb]
\centering
  \topcaption{Same as Table~\ref{tab:limits_separated_S}, but for DM mediators with pseudoscalar couplings.  Again, mediator couplings correspond to $\gq=g_{\chi}=1$.}
  \label{tab:limits_separated_PS}
\begin{tabular}{c|c|c|c|c|c|c|c|c|c|c}\hline
& \multicolumn{8}{c|}{$\mu(\ttbar + \mathrm{a} \to \ttbar \XX)$} & \multicolumn{2}{c}{$\mu(\bbbar + \mathrm{a} \to \bbbar \XX)$} \\
\cline{2-11}
                      & \multicolumn{2}{c|}{Dileptonic}                                 & \multicolumn{2}{c|}{\semileptonic}                             & \multicolumn{2}{c|}{All-hadronic}                                & \multicolumn{2}{c|}{\multirow{2}{*}{$\bbMET$}} & \multicolumn{2}{c}{\multirow{2}{*}{$\bbMET$}}                 \\
$m_{\mathrm{a}}$, $m_{\chi}$ & \multicolumn{2}{c|}{ $\ttMET$}  & \multicolumn{2}{c|}{$\ttMET$} & \multicolumn{2}{c|}{$\ttMET$} & \multicolumn{2}{c|}{} & \multicolumn{2}{c}{}                                                \\
\cline{2-11}
[\GeVns{}] & Obs. & Exp. & Obs. & Exp. & Obs. & Exp. & Obs. & Exp. & Obs. & Exp. \\
\hline
 10,   1 &  51             &  26 & 4.5 & 3.6 & 2.2 &  2.4  & 26  & 21  &     1.5$\times10^4$ & 1.2$\times10^4$   \\
 20,   1 &  55             &  26 & 3.8 & 3.0 & 2.6 &  3.1  & 42  & 35  &     141  & 117                          \\
 50,   1 &  24             &  23 & 2.9 & 2.7 & 2.5 &  3.0  & 54  & 41  &      95  &  68                          \\
 100,  1 &  38             &  29 & 3.6 & 3.7 & 2.4 &  3.3  & 60  & 37  &     116  &  81                          \\
 200,  1 &  89             &  64 & 7.0 & 6.3 & 4.4 &  4.9  & 58  & 68  &     262  & 214                          \\
 300,  1 & 133             & 123 &  11 &  10 & 5.3 &  6.9  & 105  & 95  &     625  & 611                          \\
 500,  1 & 1.0$\times10^3$ & 729 &  59 &  56 &  32 &   42  & 626 & 697 &     3.8$\times10^3$ & 4.4$\times10^3$   \\
\hline
\end{tabular}
\end{table*}

\subsection{Combined search results}\label{sec:results_combined}
Signal region yields obtained from a simultaneous background-only fit of all of the search channels are similar to those listed in Table~\ref{tab:postfit_yields_separated_PS}.  Fitted $\ptmiss$ distributions in the eight SRs are nearly indistinguishable from those of Figs.~\ref{fig:postfit_ptmiss_separated_PS_lep} and~\ref{fig:postfit_ptmiss_separated_PS_nolep}.  The nuisance parameter shifts in the combined fit are consistent with those of the individual channel fits, while the fit uncertainty in the b tagging efficiency nuisance parameter becomes more tightly constrained.  The $p$-value of the saturated likelihood goodness-of-fit test is 0.11, which indicates no significant deviation with respect to background predictions.

\begin{figure}[h!tb]
\centering
  \includegraphics[width=0.49\textwidth]{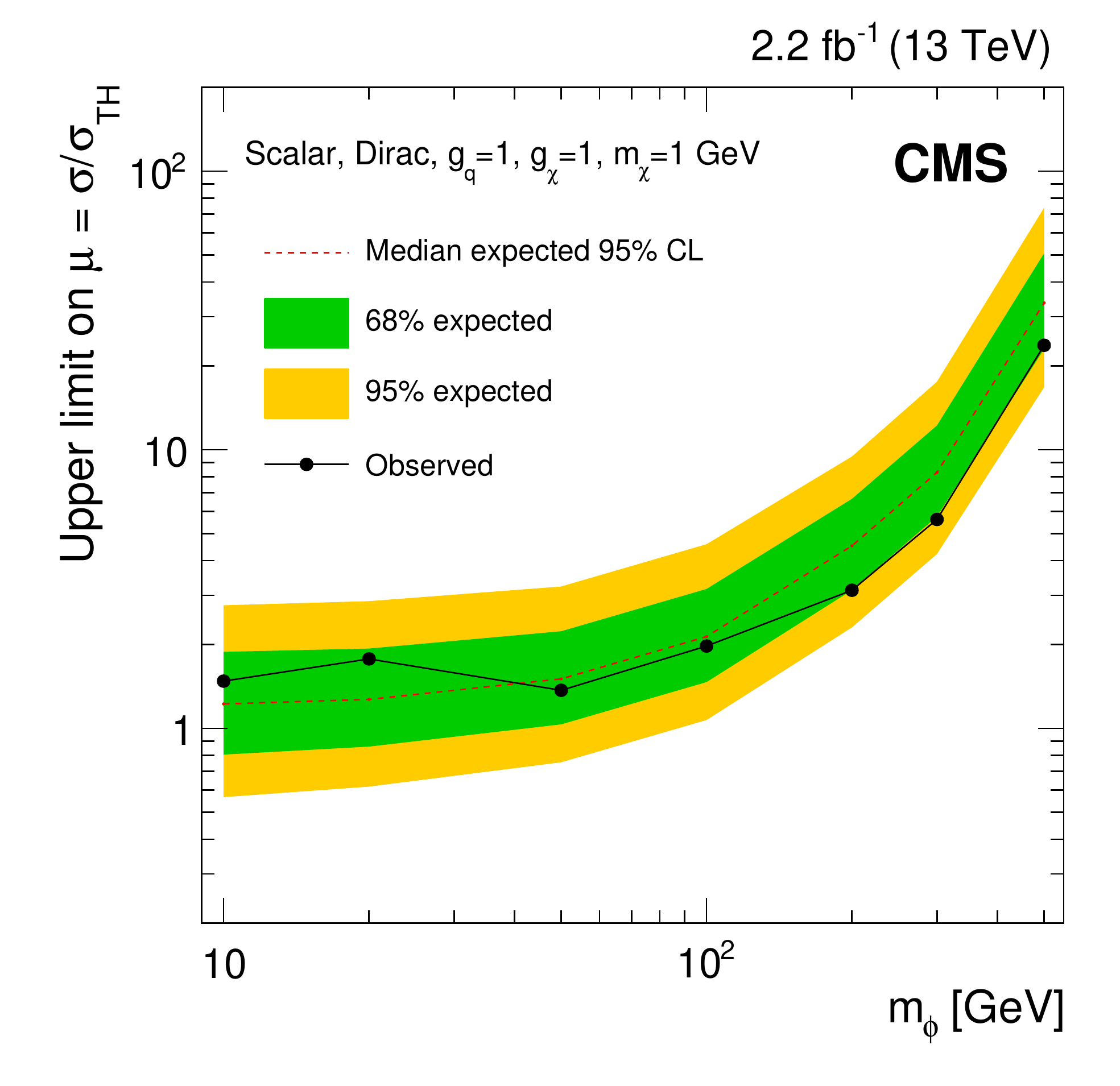}
  \includegraphics[width=0.49\textwidth]{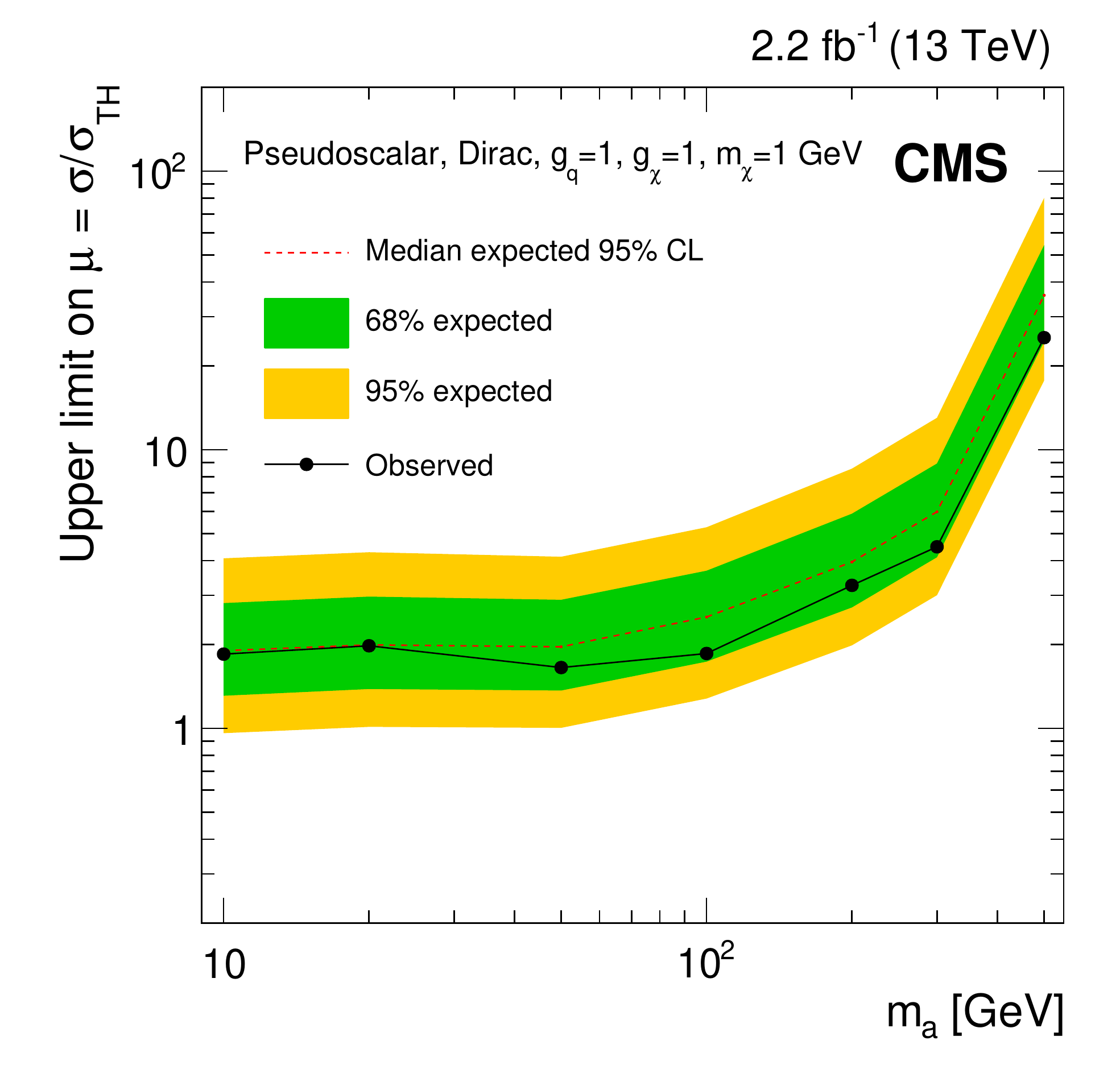}
  \caption{The ratios ($\mu$) of the 95\% CL upper limits on the combined $\ttDM$ and $\bbDM$ cross section to simplified model expectations.  The limits are obtained from combined fits to the $\ttMET$ and $\bbMET$ signal and background control regions for the hypothesis of a scalar mediator (\cmsLeft) and a pseudoscalar mediator (\cmsRight).  A fermionic DM particle with a mass of $1\GeV$ is assumed in both panels.  Mediator couplings correspond to $\gq=g_{\chi}=1$.}
  \label{fig:limits_combined}
\end{figure}

A simultaneous signal+background fit is performed using all SRs and CRs, and 95\% CL upper limits are set on the cross section ratio $\mu$ for DM produced in association with heavy-flavor quark pairs.  Table~\ref{tab:postfit_limits_combined} provides limits obtained for the scalar and pseudoscalar mediator hypotheses.  These limits are presented graphically in Fig.~\ref{fig:limits_combined}.  The combination of $\ttMET$ and $\bbMET$ search channels enhances sensitivity to both the scalar and the pseudoscalar mediator scenarios.

\begin{table*}[h!tb]
\centering
  \topcaption{Observed and expected 95\% CL upper limits on the ratio ($\mu$) of the combined $\ttDM$ and $\bbDM$ cross sections to the simplified model expectation.  The limits are obtained from a combined fit to all signal and background control regions. DM mediators with scalar or pseudoscalar couplings are assumed.  Mediator couplings correspond to $g_{q}=g_{\chi}=1$.}
  \label{tab:postfit_limits_combined}
\begin{tabular}{c|c|c|c|c|c|c}\hline
$m_{\phi/\mathrm{a}}$, $m_{\chi}$ & \multicolumn{3}{c|}{$\mu(\ttbb + \phi ~\to~ \ttbar\XX/\bbbar\XX)$} & \multicolumn{3}{c}{$\mu(\ttbb + \mathrm{a}\text{ to } \ttbar\XX/\bbbar\XX)$} \\
\cline{2-7}
[\GeVns{}]    & Obs. & Exp. & [$-$1 s.d., $+$1 s.d.] & Obs. & Exp. & [$-$1 s.d., $+$1 s.d.] \\ \hline
 10,   1 &     1.5  &  1.2   & [0.8,  1.9]  &  1.8 & 1.9 & [1.3, 2.8] \\
 20,   1 &     1.8  &  1.3   & [0.9,  1.9]  &  2.0 & 2.0 & [1.4, 3.0] \\
 50,   1 &     1.4  &  1.5   & [1.0,  2.2]  &  1.6 & 2.0 & [1.4, 2.9] \\
 100,  1 &     2.0  &  2.1   & [1.5,  3.2]  &  1.9 & 2.5 & [1.7, 3.7] \\
 200,  1 &     3.1  &  4.5   & [3.1, 6.7]   &  3.3 & 3.9 & [2.7, 5.9] \\
 300,  1 &     5.6  &  8.3   & [5.8, 12]    &  4.5 & 6.0 & [4.1, 8.9] \\
 500,  1 &     24   &  34    & [23, 51]     &	25 &  36 & [24, 54]   \\
\hline
\end{tabular}
\end{table*}

Signal cross sections may be scaled to larger values of $\gq$ and $g_{\chi}$ using the relationship given in Ref.~\cite{Abercrombie:2015wmb}.  This simple scaling approximation is valid as long as the mediator width remains below 20\% of its mass.  With $\gq=g_{\chi}=1.5$, the relative width of the 500\GeV scalar (pseudoscalar) mediator is 14\% (18\%).  The relative width decreases with decreasing mediator mass.  For coupling values of $\gq=g_{\chi}=1.5$, the $\ptmiss$ distributions of the various mediator hypotheses are also unchanged with respect to those obtained with $\gq=g_{\chi}=1$, thus the limits of Fig.~\ref{tab:postfit_limits_combined} may be scaled accordingly~\cite{Abercrombie:2015wmb}.  Assuming coupling values of $\gq=g_{\chi}=1.5$, the observed (expected) 95\% CL exclusions are $m_{\phi} < 124\,(105)\GeV$ for a scalar mediator, and $m_{\mathrm{a}} < 128~(76)\GeV$ for a pseudoscalar mediator.

\section{Summary}\label{sec:summary}
A search for an excess of events with large missing transverse momentum (\ptmiss) produced in association with a pair of heavy-flavor quarks has been performed with a sample of proton-proton interaction data at a center-of-mass energy of 13 TeV. The data correspond to an integrated luminosity of 2.2\fbinv collected with the CMS detector at the CERN LHC.  The analysis explores $\bbMET$ and the dileptonic, \semileptonic, and all-hadronic $\ttMET$ final states.  A resolved top quark tagger is used to categorize events in the all-hadronic channel.  No significant deviation from the standard model background prediction is observed.  Results are interpreted in terms of dark matter (DM) production, and constraints are placed on the parameter space of simplified models with scalar and pseudoscalar mediators.  The DM search channels are considered both individually and, for the first time, in combination.  The combined search excludes production cross sections larger than 1.5 or 1.8 times the values predicted for a 10\GeV scalar mediator or a 10\GeV pseudoscalar mediator, respectively, for couplings of $\gq=g_{\chi}=1$.  The limits presented are the first achieved on simplified models of dark matter produced in association with heavy-flavor quark pairs.

\begin{acknowledgments}
We congratulate our colleagues in the CERN accelerator departments for the excellent performance of the LHC and thank the technical and administrative staffs at CERN and at other CMS institutes for their contributions to the success of the CMS effort. In addition, we gratefully acknowledge the computing centers and personnel of the Worldwide LHC Computing Grid for delivering so effectively the computing infrastructure essential to our analyses. Finally, we acknowledge the enduring support for the construction and operation of the LHC and the CMS detector provided by the following funding agencies: BMWFW and FWF (Austria); FNRS and FWO (Belgium); CNPq, CAPES, FAPERJ, and FAPESP (Brazil); MES (Bulgaria); CERN; CAS, MoST, and NSFC (China); COLCIENCIAS (Colombia); MSES and CSF (Croatia); RPF (Cyprus); SENESCYT (Ecuador); MoER, ERC IUT, and ERDF (Estonia); Academy of Finland, MEC, and HIP (Finland); CEA and CNRS/IN2P3 (France); BMBF, DFG, and HGF (Germany); GSRT (Greece); OTKA and NIH (Hungary); DAE and DST (India); IPM (Iran); SFI (Ireland); INFN (Italy); MSIP and NRF (Republic of Korea); LAS (Lithuania); MOE and UM (Malaysia); BUAP, CINVESTAV, CONACYT, LNS, SEP, and UASLP-FAI (Mexico); MBIE (New Zealand); PAEC (Pakistan); MSHE and NSC (Poland); FCT (Portugal); JINR (Dubna); MON, RosAtom, RAS, RFBR and RAEP (Russia); MESTD (Serbia); SEIDI, CPAN, PCTI and FEDER (Spain); Swiss Funding Agencies (Switzerland); MST (Taipei); ThEPCenter, IPST, STAR, and NSTDA (Thailand); TUBITAK and TAEK (Turkey); NASU and SFFR (Ukraine); STFC (United Kingdom); DOE and NSF (USA).

\hyphenation{Rachada-pisek} Individuals have received support from the Marie-Curie program and the European Research Council and Horizon 2020 Grant, contract No. 675440 (European Union); the Leventis Foundation; the A. P. Sloan Foundation; the Alexander von Humboldt Foundation; the Belgian Federal Science Policy Office; the Fonds pour la Formation \`a la Recherche dans l'Industrie et dans l'Agriculture (FRIA-Belgium); the Agentschap voor Innovatie door Wetenschap en Technologie (IWT-Belgium); the Ministry of Education, Youth and Sports (MEYS) of the Czech Republic; the Council of Science and Industrial Research, India; the HOMING PLUS program of the Foundation for Polish Science, cofinanced from European Union, Regional Development Fund, the Mobility Plus program of the Ministry of Science and Higher Education, the National Science Center (Poland), contracts Harmonia 2014/14/M/ST2/00428, Opus 2014/13/B/ST2/02543, 2014/15/B/ST2/03998, and 2015/19/B/ST2/02861, Sonata-bis 2012/07/E/ST2/01406; the National Priorities Research Program by Qatar National Research Fund; the Programa Clar\'in-COFUND del Principado de Asturias; the Thalis and Aristeia programs cofinanced by EU-ESF and the Greek NSRF; the Rachadapisek Sompot Fund for Postdoctoral Fellowship, Chulalongkorn University and the Chulalongkorn Academic into Its 2nd Century Project Advancement Project (Thailand); and the Welch Foundation, contract C-1845.
\end{acknowledgments}

\bibliography{auto_generated}

\cleardoublepage \appendix\section{The CMS Collaboration \label{app:collab}}\begin{sloppypar}\hyphenpenalty=5000\widowpenalty=500\clubpenalty=5000\textbf{Yerevan Physics Institute,  Yerevan,  Armenia}\\*[0pt]
A.M.~Sirunyan, A.~Tumasyan
\vskip\cmsinstskip
\textbf{Institut f\"{u}r Hochenergiephysik,  Wien,  Austria}\\*[0pt]
W.~Adam, E.~Asilar, T.~Bergauer, J.~Brandstetter, E.~Brondolin, M.~Dragicevic, J.~Er\"{o}, M.~Flechl, M.~Friedl, R.~Fr\"{u}hwirth\cmsAuthorMark{1}, V.M.~Ghete, C.~Hartl, N.~H\"{o}rmann, J.~Hrubec, M.~Jeitler\cmsAuthorMark{1}, A.~K\"{o}nig, I.~Kr\"{a}tschmer, D.~Liko, T.~Matsushita, I.~Mikulec, D.~Rabady, N.~Rad, B.~Rahbaran, H.~Rohringer, J.~Schieck\cmsAuthorMark{1}, J.~Strauss, W.~Waltenberger, C.-E.~Wulz\cmsAuthorMark{1}
% ...
\vskip\cmsinstskip
1:~~Also at Vienna University of Technology, Vienna, Austria\\
% ...

\end{sloppypar}
\end{document}